\documentclass[11pt]{article}
\pdfoutput=1

\usepackage[usenames,dvipsnames]{xcolor}
\usepackage[a4paper]{geometry}
\usepackage[english]{babel}
\usepackage{cite,enumerate,enumitem,booktabs,float,graphicx}

\usepackage[T1]{fontenc}
\usepackage[utf8]{inputenc}
\usepackage{lmodern}

\makeatletter
\g@addto@macro\bfseries{\boldmath}
\makeatother

\usepackage{bm,amsmath,amssymb,tensor,mathtools}
\numberwithin{equation}{section}
\usepackage{nth}

\usepackage{tikz}
\usetikzlibrary{cd}

\usepackage{caption}
\usepackage{subcaption}

\usepackage[pdftex]{hyperref}
\definecolor{dark-blue}{rgb}{0.15,0.15,0.4}
\hypersetup{
  colorlinks,
  linkcolor={dark-blue},
  citecolor={blue}
}

\newcommand{\RR}{\mathbb{R}}
\newcommand{\CC}{\mathbb{C}}
\newcommand{\PP}{\mathbb{P}}

\newcommand{\pd}{\partial}

\newcommand{\vphi}{\varphi}

\newcommand{\inv}{^{-1}}
\newcommand{\OO}{\mathcal{O}}
\newcommand{\qiq}{\quad\implies\quad}

\def\mN{\mathcal{N}}

\usepackage{enumitem}
\usepackage{amsthm}

\usepackage{authblk}

\title{\textbf{%
  Spin Matrix Theory String Backgrounds
  \\
  and Penrose Limits of AdS/CFT
}
}
\date{}
\author[1]{Troels Harmark}
\author[2]{Jelle Hartong}
\author[1,3]{Niels A. Obers}
\author[1]{Gerben Oling}
\affil[1]{%
  The Niels Bohr Institute,
  University of Copenhagen,
  \protect\\
  Blegdamsvej 17,
  DK-2100 Copenhagen Ø,
  Denmark
}
\affil[2]{%
  School of Mathematics and Maxwell Institute for Mathematical Sciences,
  \protect\\
  University of Edinburgh,
  Peter Guthrie Tait Road,
  Edinburgh EH9 3FD, UK
}
\affil[3]{%
  Nordita,
  KTH Royal Institute of Technology and Stockholm University,
  \protect\\
  Roslagstullsbacken 23,
  SE-106 91 Stockholm,
  Sweden
}

\begin{document}

\maketitle
\thispagestyle{empty}

\begin{abstract}
\noindent
Spin Matrix theory (SMT) limits provide a way to capture the dynamics of the AdS/CFT correspondence near BPS bounds.
On the string theory side, these limits result in non-relativistic sigma models that can be interpreted as novel non-relativistic strings.
This SMT string theory couples to non-relativistic $U(1)$-Galilean background geometries.
In this paper, we explore the relation between pp-wave backgrounds obtained from Penrose limits of AdS${}_5 \times S^5$, and a new type of $U(1)$-Galilean backgrounds that we call flat-fluxed (FF) backgrounds.
These FF backgrounds are the simplest possible SMT string backgrounds and correspond to free magnons from the spin chain perspective.
We provide a catalogue of the $U(1)$-Galilean backgrounds one obtains from SMT limits of string theory on AdS${}_5 \times S^5$ and subsequently study large charge limits of these geometries from which the FF backgrounds emerge.
We show that these limits are analogous to Penrose limits of AdS${}_5 \times S^5$ and demonstrate that the large charge/Penrose limits commute with the SMT limits.
Finally, we point out that $U(1)$-Galilean backgrounds prescribe a symplectic manifold for the transverse SMT string embedding fields.
This is illustrated with a Hamiltonian derivation for the SMT limit of a particle.
\end{abstract}

\newpage
\tableofcontents

\section{Introduction}
\label{sec:intro}
Holographic dualities are central to our understanding
of the quantum properties of space-time, such as microscopic explanations of black hole entropy.
The concrete realization of holography in terms of the AdS/CFT correspondence \cite{Maldacena:1997re} provides, in its strongest form,
a quantitative agreement between $SU(N)$ ${\cal{N}} = 4$ super-Yang--Mills theory (SYM) and type IIB string theory on AdS${}_5 \times S^5$ for any $N$ and any value of the 't~Hooft coupling $\lambda$.
Enormous progress has been made in the supersymmetric sector using localization techniques \cite{Pestun:2007rz} as well as the planar $N\rightarrow \infty$ limit due to the link to integrable spin chains \cite{Minahan:2002ve,Beisert:2003tq}.
However, although this may be challenging, obtaining a complete understanding of black holes in AdS/CFT inevitably requires us to venture beyond both the supersymmetric and large $N$ realms.

One path in this direction is provided by novel tractable limits of the AdS/CFT correspondence.
This approach includes in particular Spin Matrix theory (SMT), which describes near-BPS limits of AdS${}_5$/CFT${}_4$, as proposed in~\cite{Harmark:2014mpa}.
In further detail, SMT is a quantum-mechanical theory described by a Hamiltonian consisting of harmonic oscillator operators that transform both
in the adjoint (matrix) representation of $SU(N)$,
as well as
in a particular `spin' subgroup~$G_s$ of the global superconformal $PSU(2,2|4)$ symmetries of $\mN=4$ SYM.
The precise form of the Hamiltonian, and in particular the relevant spin group $G_s$, depends on the details of the BPS bound that is considered, giving rise to a set of possible SMTs.
In addition, an alternative formulation of one such theory (associated to a particular spin group)  as a simple two-dimensional non-relativistic field theory was recently obtained~\cite{Harmark:2019zkn} and has since been generalized to other spin groups \cite{Baiguera:2020jgy}.

In parallel to these developments on the field theory side, it was first realized in \cite{Harmark:2017rpg} and subsequently elaborated in \cite{Harmark:2018cdl,Harmark:2019upf} that, from a bulk perspective, SMT takes the form of a non-relativistic string theory with a particular type of non-relativistic target space known as $U(1)$-Galilean geometry.%
\footnote{%
  The name $U(1)$-Galilean refers to the fact that the geometry can be obtained~\cite{Harmark:2017rpg} from the gauging of the direct sum of the Galilean algebra and $U(1)$, similar to how torsional Newton--Cartan geometry can be obtained by gauging the Bargmann algebra~\cite{Andringa:2010it,Hartong:2015zia}.
  The former algebra can be obtained from a contraction of the latter.
  The development of strings on Newton--Cartan backgrounds is reviewed below.
}
This strongly suggests that the dual Spin Matrix theories are associated to an emergent non-relativistic geometry in the bulk.
The appearance of non-relativistic symmetries is expected from
the dual theory, since zooming in on the unitarity bounds of ${\cal{N}}=4$ SYM on $R \times S^3$ may be thought of as a non-relativistic limit \cite{Harmark:2014mpa}.
This can be seen most directly from the fact that the relativistic magnon dispersion relation \cite{Beisert:2005tm} of the ${\cal{N}}=4$ spin chain exhibits non-relativistic features in the SMT limit \cite{Harmark:2008gm}.
As such, SMT provides us with a concrete and well-defined framework that allows us to formulate a holographic correspondence involving non-relativistic string theory and its non-relativistic background geometries.

These advances in our understanding of SMT strings are part of a broader development of non-relativistic string theory in recent years~\cite{Andringa:2012uz,
  Harmark:2017rpg,Kluson:2018egd,Bergshoeff:2018yvt,Kluson:2018grx,Harmark:2018cdl,
  Kluson:2019ifd,Gomis:2019zyu,Gallegos:2019icg,Roychowdhury:2019olt,Bergshoeff:2019pij,
  Blair:2019qwi,Yan:2019xsf,Gomis:2020fui,Roychowdhury:2020dke,Roychowdhury:2020yun}, spurred on in part by new insights into
Newton-Cartan (NC) geometry,
including the discovery of torsional Newton-Cartan (TNC) geometry \cite{Christensen:2013lma,Christensen:2013rfa,Hartong:2014pma} and stringy Newton-Cartan (SNC) geometry
\cite{Andringa:2012uz}.
Such NC-type geometries allow for a covariant formulation of physics in non-relativistic limits, $1/c$ expansions and reductions.
Furthermore, these notions of geometry allow one to generalize the flat space Gomis--Ooguri non-relativistic string action~\cite{Gomis:2000bd,Danielsson:2000gi} to arbitrary backgrounds.
The resulting non-relativistic string theory is a unitary UV-completion of (a particular version of) non-relativistic gravity,%
\footnote{%
  See~\cite{VandenBleeken:2017rij,Hansen:2018ofj,Hansen:2020pqs,Ergen:2020yop}
  for the non-relativistic expansion of general relativity.
}
and quantum consistency conditions \cite{Gomis:2019zyu,Gallegos:2019icg,Yan:2019xsf} that describe the dynamics of this gravity theory have recently been obtained,
both for the TNC string  \cite{Harmark:2017rpg,Harmark:2018cdl} as well as the SNC string \cite{Andringa:2012uz,Bergshoeff:2018yvt,Yan:2019xsf}.
Moreover, a map between the actions of non-relativistic string theory with TNC%
\footnote{%
  More precisely, the target space geometry of this string theory is TNC geometry
plus an extra periodic target space direction where there is a non-zero string winding.
  Classically, the embedding coordinate along this direction is locally pure gauge, but this is not true globally due to the winding.
}
and SNC target spacetimes has been constructed in Ref.~\cite{Harmark:2019upf}.
In all approaches, an important motivation for studying this non-relativistic corner of string theory is that it may teach us lessons that are ultimately relevant for relativistic quantum gravity/string theory.

The SMT string, which is the main subject of the present paper, was originally derived from the TNC string
by performing the stringy equivalent of the SMT limit~\cite{Harmark:2017rpg,Harmark:2018cdl}.
While TNC/SNC strings are two-dimensional relativistic conformal field theories on the worldsheet, this limit results in a non-relativistic sigma model with scaling symmetry.
In particular, SMT string theory has a non-relativistic version of Weyl symmetry, and the corresponding reparametrization symmetries are given by the Galilean conformal algebra.
Furthermore, as mentioned above,
the backgrounds that SMT strings couple to is a variation on NC geometry which is known as $U(1)$-Galilean geometry.

Currently, the most important open questions surrounding SMT strings concern their quantization and their quantum consistency conditions, which would provide the dynamics of $U(1)$-Galilean geometry.
As a first step towards understanding the quantum theory, it is important to understand the simplest non-trivial backgrounds of these strings, i.e.\ the analogue of Minkowski space for the relativistic string.
This is non-trivial since
it turns out that the theory has no dynamics on a completely flat $U(1)$-Galilean geometry
(which is essentially flat NC space plus
a vanishing $U(1)$ gauge connection)
since the gauge-fixed action on this background does not contain any time derivatives.

\begin{figure}[t]
  \centering
  \includegraphics[width=0.9\linewidth]{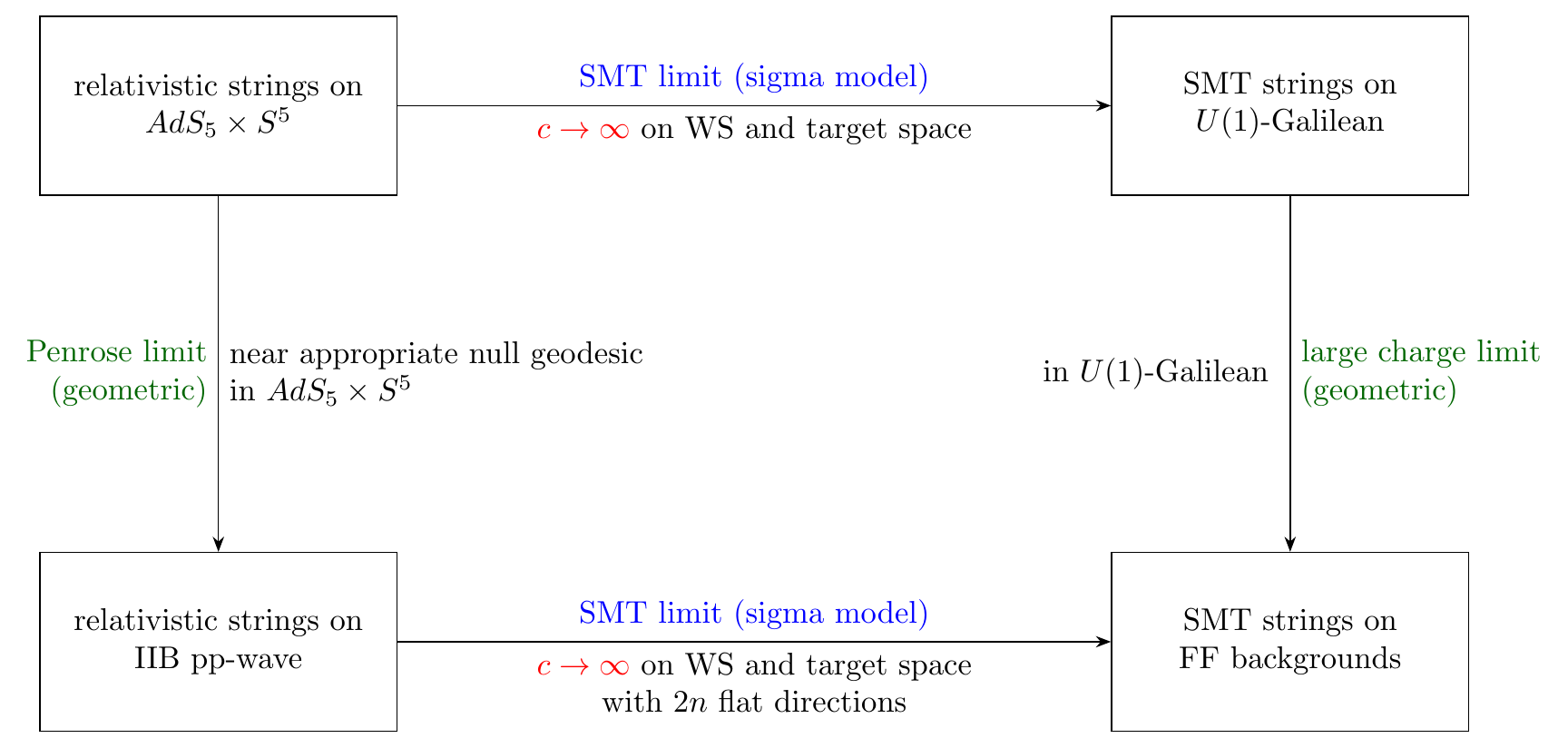}
  \caption{Overview of SMT and large charge/Penrose limits.}%
  \label{fig:smt-penrose-limits-diagram}
\end{figure}

In this paper we will construct a class of backgrounds for SMT strings, which we will refer to as flat-fluxed (FF) backgrounds,
that can be considered the analogue of Minkowski space.
Crucially, on these backgrounds the non-relativistic sigma-model reduces to a free theory, at least in the equivalent of `light-cone' gauge.
Thus the resulting worldsheet theories
can be considered to be the simplest in the same sense that the Polyakov action on Minkowski target space
is the simplest relativistic string theory.

The FF backgrounds arise from curved backgrounds in a large charge limit, which is analogous to the Penrose limit on the AdS side \cite{Blau:2002dy}.
Since the worldsheet theories are expected to be difficult to quantize on general curved SMT backgrounds, this large charge limit is convenient because the resulting geometry is (mostly) flat.
Furthermore, we show that the SMT and large charge/Penrose limits commute, as illustrated in Figure~\ref{fig:smt-penrose-limits-diagram}.

We obtain the curved $U(1)$-Galilean backgrounds from the Lorentzian AdS${}_5 \times S^5$ type IIB background by implementing the SMT limit on the string sigma model.
The particular properties of these backgrounds depend on the particular SMT limit, which determines the specific limit of AdS$_5\times S^5$ that we consider, in direct analogy to how the spin group $G_s$ and other properties of the field theory dual depend on the form of the near-BPS limit under consideration~\cite{Harmark:2008gm,Harmark:2014mpa}.

In this paper, we provide a general prescription that allows us to obtain the $U(1)$-Galilean and FF backgrounds for all integer SMT limits on AdS$_5\times S^5$.
We will illustrate this prescription with three representative examples.
The first involves the largest possible compact spin group $SU(2|3)$, which originates from the SMT limit with all three R-charges on the $S^5$ in the BPS bound.
The second has the non-compact spin group $SU(1,1)$ and involves both a non-zero R-charge on the $S^5$ and one angular momentum on the AdS${}_5$ factor.
Finally, we consider the most general SMT limit, which gives a theory with spin group $PSU(1,2|3)$.
The latter theory contains all other SMTs and we illustrate how their associated backgrounds can be recovered from the $PSU(1,2|3)$ background.
For each of these examples we first find the curved SMT $U(1)$-Galilean backgrounds and subsequently take a large charge limit of these geometries.
We then show that the resulting FF backgrounds can also be obtained by first taking an appropriate Penrose limit of AdS$_5\times S^5$ and subsequently taking the SMT limit.

When performing this limit for each of the SMT spin groups, one obtains a particular $(2n+1)$-dimensional FF geometry
(with $n=1,2,4)$ consisting of an exact clock one-form, a flat spatial metric on equal time slices and a non-zero flux of the $U(1)$ connection in the $2n$ spatial
dimensions.
A general FF background consists of $n$ two-planes, each with an associated $U(1)$ flux.
We thus arrive at a particular class of $U(1)$-Galilean backgrounds that is directly related to AdS/CFT via the
SMT and large charge/Penrose limits, giving rise to worldsheet theories that are expected to be well-behaved quantum theories.

For the simplest spin group $SU(2)$ (corresponding to two non-zero R-charges
in the BPS bound), the resulting worldsheet theory is the well-known $SU(2)$ Landau-Lifshitz model.%
\footnote{%
  This connection follows since the SMT limit is closely related to limits of spin chains
\cite{Kruczenski:2003gt}
  (see also  \cite{Kruczenski:2004kw,Hernandez:2004uw,Stefanski:2004cw,Bellucci:2004qr,Bellucci:2006bv})
  that have been studied in connection to integrability in AdS/CFT.
In particular, for the $SU(2)$ case we obtain a covariant version
 \cite{Harmark:2017rpg,Harmark:2018cdl}  of the Landau-Lifshitz sigma-model which is the continuum limit of the ferromagnetic $XXX_{1/2}$ Heisenberg spin chain.
 }
After the flat space limit, this means that the theory on the $SU(2)$ FF background is the free limit of the Landau-Lifshitz model, i.e.\ a theory of a free magnon.
In this case $n=1$ and there is only one parameter, namely the flux on $\RR^2$, corresponding to the mass of the magnon.
More generally, this means that the theory on a given $(2n+1)$-dimensional FF background can be interpreted  as $n$ free magnons, where the particular value of the flux in each of the $n$ planes corresponds to the mass of the respective magnon.
We will elaborate more on the connection between our results and earlier work on integrable sigma models arising from long wavelength limits of spin chains at the end of this paper.

Finally, we formulate a new and perhaps surprising interpretation of the $U(1)$-Galilean backgrounds of SMT string theory.
In the flat worldsheet gauge-fixed SMT action, the velocities of the embedding fields appear only linearly.
The form of the action then suggest that one can see the $U(1)$ background flux as a potential for the symplectic form on a phase space parametrized by the embedding fields after gauge fixing.
This interpretation is verified explicitly by a detailed analysis of the SMT limit of a point particle propagating on a pp-wave background of the type appearing in the lower left corner of the diagram in Figure~\ref{fig:smt-penrose-limits-diagram}.

A brief outline of the paper is as follows.
In Section~\ref{sec:smt-review} we first review the main elements of the SMT limit on both the field theory and string theory side.
We describe the near-BPS limits of ${\cal{N}} = 4$  super-Yang--Mills that are used to construct SMT and present the possible spin groups that follow from zooming in on integer BPS bounds.
We then show how this limiting procedure is translated to the string theory side as a limit of relativistic strings on AdS${}_5 \times S^5$, and we briefly comment on how this construction is related to strings on TNC backgrounds.
The main points of this section are the notion of $U(1)$-Galilean geometry and the corresponding flat worldsheet gauge-fixed SMT action~\eqref{eq:smt-flat-gf-string-action} coupling to this background geometry.

We then proceed in Section~\ref{sec:smt-string-backgrounds} by constructing all curved $U(1)$-Galilean geometries associated to SMT limits.
We provide a systematic procedure to obtain these backgrounds for a given SMT spin group and illustrate this for the cases $SU(2|3)$ and $SU(1,1)$, which have not appeared in the literature before.
Importantly, the construction involves a coordinate system that allows for a suitable spin chain interpretation of the resulting sigma model.
We also derive the full $PSU(1,2|3)$ geometry in these coordinates, and we show how the geometries associated to the smaller spin groups can be obtained as submanifolds of this geometry.

In Section~\ref{sec:penrose-limits} we present one of our main results, namely the FF backgrounds that represent the simplest non-trivial target spaces of the SMT string.
For these, the SMT string action~\eqref{eq:smt-flat-gf-string-action} leads to a non-relativistic worldsheet model that corresponds to free magnons from the perspective of spin chains.
In this section, we also show that the SMT and large charge/Penrose limits commute, as illustrated in Figure~\ref{fig:smt-penrose-limits-diagram}.
This is demonstrated explicitly for the $SU(2|3)$, $SU(1,1)$ and $PSU(1,2|3)$ geometries, and all other cases can easily be shown similarly.

Next, Section~\ref{sec:point-particle-limit} puts forward the new interpretation of the $U(1)$-Galilean background geometry of the SMT string as providing the symplectic geometry of the phase space of the embedding fields.
We will show in detail how this interpretation arises from the SMT limit of a point particle
on a pp-wave background of the type that is encountered in the previous section.
This interpretation will be further examined in an upcoming work on the Hamiltonian analysis of SMT strings~\cite{upcoming-hamiltonian-paper}.

We end the paper in Section~\ref{sec:outlook} with a discussion of our results, and put them into context with earlier results on integrable sigma-models arising from spin chains and their connection to AdS/CFT.
Finally, we list future lines of investigation and open problems.

\section{Brief review of Spin Matrix limits of fields and strings}
\label{sec:smt-review}
In this section, we briefly review the limits of $\mathcal{N}=4$ super-Yang--Mills that are used to construct Spin Matrix theories~\cite{Harmark:2014mpa}.
We also review the Polyakov-type string sigma model that has been put forward to describe the bulk dual to these theories, and we discuss their proposed identification~\cite{Harmark:2017rpg,Harmark:2018cdl}.

\subsection{Field theory limit}
\label{ssec:smt-review-ft}
Spin Matrix theories (SMTs) are quantum-mechanical theories that describe the dynamics of a subsector of $SU(N)$ $\mathcal{N}=4$ super-Yang--Mills theory on $\RR\times S^3$ in terms of a Hamiltonian consisting of harmonic oscillator operators that transform both in the adjoint (matrix) representation of $SU(N)$ as well as a particular `spin' subgroup $G_s$ of $PSU(2,2|4)$.
These theories are obtained from $\mN=4$ super-Yang--Mills by zooming in on a particular BPS bound $E\geq Q$ using~\cite{Harmark:2014mpa}
\begin{equation}
  \label{eq:smt-limit-field-theory}
  \lambda \to 0,
  \qquad
  N = \text{fixed},
  \qquad
  \frac{E-Q}{\lambda} = \text{fixed}.
\end{equation}
Here, $\lambda$ is the 't Hooft coupling and $Q$
is a particular combination of
the two commuting angular momenta $S_1$ and $S_2$ coming from the $S^3$
and the three commuting R-charges $J_1$, $J_2$ and $J_3$,
which we denote by
\begin{equation}
  \label{eq:generic-charge-decomposition}
  Q = S + J,
  \qquad
  S = a^i S_i,
  \quad
  J = b^j J_j.
\end{equation}
In this work, we consider all SMT limits corresponding to BPS bounds with integer coefficients, which are listed in Table~\ref{tab:smt-limits} below.%
\footnote{%
  Note that we cannot consider the combination $S_1+S_2+J_1+J_2$ since it does not correspond to a BPS bound of the $psu(2,2|4)$ algebra.
  It may appear that the bulk construction that we present in the following allows for a SMT bulk background corresponding to this combination of charges, but this is only because we focus on the bosonic (NS-NS) couplings.
  In the bulk, the fact that no BPS bound associated to this combination exists is reflected in the fact that one can construct fermionic string states with $J_3=0$ and $S_1=S_2=J_1=J_2=1/2$, so that $E=\Delta_\text{fermion}=3/2 < S_1+S_2+J_1+J_2$.
}

\begin{table}[ht]
  \centering
  \begin{tabular}{r|l}
    Spin group $G_s$
    & Charge $Q$
    \\
    \hline
    $SU(2)$
    &
    $J_1+J_2$
    \\
    $SU(2|3)$
    &
    $J_1+J_2+J_3$
    \\
    $SU(1,1)$
    &
    $S_1 + J_1$
    \\
    $PSU(1,1|2)$
    &
    $S_1+J_1+J_2$
    \\
    $SU(1,2|2)$
    &
    $S_1+S_2+J_1$
    \\
    $PSU(1,2|3)$
    &
    $S_1+S_2+J_1+J_2+J_3$
    \\
  \end{tabular}
  \caption{Spin groups $G_s$ for the SMTs obtained from the integer BPS bounds $E\geq Q$.}
  \label{tab:smt-limits}
\end{table}

For $N\to\infty$, the resulting Spin Matrix theory is described by a particular nearest-neighbor spin chain Hamiltonian describing spin chains with length $J$.
This reflects the fact that operators charged under $S$ are represented as derivatives in the spin chain picture, so they should not be counted towards the total spin chain length.
For example, the case of $Q=J=J_1+J_2$ leads to a Heisenberg spin chain, which results in the $SU(2)$ Landau-Lifshitz model in the continuum limit.
Taking $1/N$ corrections into account allows for splitting and joining of the spin chains.

\subsection{Spin Matrix strings}
\label{ssec:smt-review-strings}
In previous work~\cite{Harmark:2017rpg,Harmark:2018cdl} a dual version of this limiting procedure was developed for
relativistic string theory on AdS$_5\times S^5$,
where one takes the limit
\begin{equation}
  \label{eq:smt-limit-strings}
  g_s \to 0,
  \qquad
  N = \text{fixed},
  \qquad
  \frac{E-Q}{g_s} = \text{fixed},
\end{equation}
in direct analogy with the field theory limit~\eqref{eq:smt-limit-field-theory}.
Here, $g_s$ is the string coupling and $N$ counts the units of five-form flux, which are related to the 't Hooft coupling by $\lambda = 4\pi g_s N$.
The charges $E$ and $Q$ refer to conserved charges of the string corresponding to the isometries of AdS$_5\times S^5$.
In terms of the global coordinates
\begin{equation}
  ds^2 = R^2 \left[
    -\cosh^2\rho\, dt^2 + d\rho^2 + \sinh^2\rho\, d\Omega_3^2 + d\Omega_5^2
  \right],
\end{equation}
we have $E = i\pd_t$.
Furthermore, if we have $Q=S+J$ as in~\eqref{eq:generic-charge-decomposition}
we can choose angular coordinates
$\bar\gamma$ on the $S^3\subset\text{AdS}_5$ and $\gamma$ on the $S^5$ so that
$S = -i\pd_{\bar\gamma}$ and $J=-i\pd_\gamma$.
To consider the limit~\eqref{eq:smt-limit-strings}, it is then useful to introduce coordinates $x^0$ and $u$ such that%
\footnote{\label{fn:first-u-footnote}%
  Note that in previous discussions~\cite{Harmark:2018cdl} of some of the backgrounds we will consider here, an alternative coordinate choice was used where $-i\pd_u=(E+Q)/2$.
  The resulting geometries are less natural from the spin chain perspective when angular momenta on AdS$_5$ are involved,
  and the present coordinate choice moreover has the advantage that the corresponding gauge-fixed string action takes a much simpler form.
}
\begin{equation}
  \label{eq:x0-u-vectors-charges}
  i\pd_{x^0}
  = E - Q
  = E - S - J,
  \qquad
  -i\pd_u = \frac{1}{2} \left(E- S + J\right).
\end{equation}
We then rescale the coordinate $x^0$ in such a way that its conserved charge scales as $g_s$ in the limit $g_s\to0$.
For this, we introduce
\begin{equation}
  \label{eq:x0-rescaling-gs}
  x^0 = \frac{\tilde{x}^0}{4\pi g_s N}.
\end{equation}
so that we can also write the SMT limit~\eqref{eq:smt-limit-strings} as
\begin{equation}
  \label{eq:smt-limit-strings-using-c}
  c\to\infty,
  \qquad
  x^0 = c^2 \tilde{x}^0,
  \quad
  c = \frac{1}{\sqrt{4\pi g_s N}},
  \quad
  N \text{ and } \tilde{x}^0 \text{ fixed}.
\end{equation}
This corresponds to a non-relativistic limit on the string worldsheet~\cite{Harmark:2017rpg,Harmark:2018cdl}.
Note that the momentum $-i\pd_u$ is given by $J$ in the SMT limit~\eqref{eq:smt-limit-strings-using-c}.

\subsubsection{Relation to non-relativistic strings}
\label{sssec:relation-to-tnc-strings}
The string theory that results from the bulk dual of the SMT limit discussed above is non-relativistic both in terms of its worldsheet geometry and its background fields.
To describe this theory and its background couplings, it is useful to first discuss another type of string that is still relativistic on the worldsheet but which couples to a non-relativistic torsional Newton--Cartan (TNC) geometry.

In Section~\ref{ssec:general-procedure-smt-string-backgrounds}, we will develop our previous discussion into a general procedure which translates the combinations of $\mN=4$ charges in Table~\ref{tab:smt-limits} to coordinate choices on AdS$_5\times S^5$.
In the SMT limit associated to these charges, the dynamics of relativistic strings on AdS$_5\times S^5$ is reduced to a particular submanifold which has a null isometry.
This submanifold can therefore be described in terms of a TNC geometry, as we briefly review below.
For this reason, TNC geometry will be a useful tool in describing the backgrounds associated to SMT limits.

Without loss of generality, one can parametrize a $(d+1)$-dimensional Lorentzian geometry with null isometry $\pd_u$ as
\begin{equation}
  \label{eq:riemannian-null-decomposition}
  ds^2 = 2\tau_\mu dx^\mu \left(du - m_\mu dx^\mu\right) + h_{\mu\nu} dx^\mu dx^\nu.
\end{equation}
Null reduction along $u$ then results in a TNC geometry, which is described by a clock one-form $\tau_\mu$, a symmetric tensor $h_{\mu\nu}$ of rank $d-1$ and a $U(1)$ connection $m_\mu$.
This decomposition is not unique, which is manifested in the gauge transformations
\begin{equation}
  \label{eq:tnc-local-transformations}
  \delta \tau_\mu = 0,
  \qquad
  \delta m_\mu = \pd_\mu \sigma + \lambda_a E^a_\mu,
  \qquad
  \delta h_{\mu\nu} = 2\tau_{(\mu} E_{\nu)}^a \lambda_a,
\end{equation}
corresponding to Galilean boosts and $U(1)$ transformations that are generated by $\lambda_a(x^\mu)$ and $\delta u = -\sigma(x^\mu)$, respectively.
Here, we have $a=1,\ldots,d$ and the $E^a_\mu$ are (spatial) vielbeine for $h_{\mu\nu} = \delta_{ab} E^a_\mu E^b_\nu$.
We can define projective inverses $h^{\mu\nu}$ and $v^\mu$ that satisfy the following completeness and orthogonality relations,
\begin{equation}
  \delta_\mu^\nu = - \tau_\mu v^\nu + h_{\mu\rho} h^{\rho\nu},
  \qquad
  v^\mu \tau_\mu = -1,
  \quad
  h^{\mu\nu}\tau_\nu = 0,
  \quad
  h_{\mu\nu}v^\nu = 0.
\end{equation}
Under the local $U(1)$ gauge transformations and Galilean boosts, we have
\begin{equation}
  \label{eq:tnc-inverses-local-transformations}
  \delta v^\mu = E^\mu_a \lambda^a,
  \qquad
  \delta h^{\mu\nu} = 0,
\end{equation}
where $E_a^\mu$ are the (projective) inverses for $E^a_\mu$.
Note that only $\tau_\mu$ and $h^{\mu\nu}$ are invariant under both Galilean boosts and $U(1)$ gauge transformations.

Strings coupling to a TNC background $(\tau_\mu,h_{\mu\nu},m_\mu)$ can be obtained starting from relativistic strings coupling to a Lorentzian metric of the form~\eqref{eq:riemannian-null-decomposition}
following the construction in~\cite{Harmark:2017rpg,Harmark:2018cdl}.
There, the (constant) momentum $P_u$ is exchanged for a single winding mode in a direction $\eta$ dual to $u$.
The resulting string theory generalizes the Gomis--Ooguri non-relativistic string~\cite{Gomis:2000bd,Danielsson:2000gi} and (at least on a classical level) can be related~\cite{Harmark:2019upf} to a gauge fixing of the `stringy Newton--Cartan' (SNC) strings~\cite{Bergshoeff:2018yvt,Bergshoeff:2019pij}.
The fact that non-relativistic geometries arise can be understood particularly nicely from the perspective of double field theory~\cite{Ko:2015rha,Morand:2017fnv}, where this construction can be interpreted in terms of a transformation similar to T-duality but along a null direction.

For most BPS bounds, the coordinate $u$ that is defined in Equation~\eqref{eq:x0-u-vectors-charges} is not null on all of AdS$_5\times S^5$.
However, as we will see in Section~\ref{ssec:general-procedure-smt-string-backgrounds}, the Spin Matrix limit confines the dynamics of the string to the submanifold $M$ where $u$ is null,
and hence
we can write the metric on this submanifold as in Equation~\eqref{eq:riemannian-null-decomposition}.
If we parametrize AdS$_5\times S^5$ using $x^0$ and $u$ from Equation~\eqref{eq:x0-u-vectors-charges}, supplemented with eight transverse coordinates $x^I$,
the nowhere-vanishing clock one-form $\tau_\mu$ must contain a component along $x^0$, so that
\begin{equation}
  \tau_\mu dx^\mu = \tau_0 dx^0 + \tau_I dx^I
  = c^2\tau_0 d\tilde{x}^0 + \OO( g_s^0 ),
  \qquad
  \tau_0 \neq 0.
\end{equation}
In the SMT limit~\eqref{eq:smt-limit-strings-using-c}, the string couples only to the leading term $\tilde\tau = \tau_0 dx^0$.
In contrast, $m_\mu$ and $h_{\mu\nu}$ are unaffected by the SMT limit, so in order to use the convenient parametrization~\eqref{eq:smt-limit-strings-using-c} of the limit we use the Galilean boosts and $U(1)$ transformations to ensure that $m_\mu$ and $h_{\mu\nu}$ do not have components along $x^0$.
Finally, in the SMT limit we also need to rescale the boost parameter $\lambda_a = \tilde{\lambda}_a/c^2$ so that the local transformations~\eqref{eq:tnc-local-transformations} reduce to
\begin{equation}
  \label{eq:u1gal-local-transformations}
  \delta \tilde\tau_\mu = 0,
  \qquad
  \delta m_\mu = \pd_\mu \sigma,
  \qquad
  \delta h_{\mu\nu} = 2\tilde\tau_{(\mu} E_{\nu)}^a \lambda_a.
\end{equation}
Note that the $U(1)$ gauge field $m_\mu$ only transforms under the $U(1)$ gauge transformations and no longer under the Galilean boosts $\lambda_a$ as in the TNC transformations~\eqref{eq:tnc-local-transformations}.
As a result, the $U(1)$ gauge field no longer mixes with the clock one-form $\tilde\tau_\mu$ and the symmetric spatial tensor $h_{\mu\nu}$.
Instead, we obtain a Galilean background $(\tau_\mu,h_{\mu\nu})$ supplemented with an independent $U(1)$ gauge field $m_\mu$.
For this reason, the geometry defined by $(\tilde\tau_\mu,h_{\mu\nu},m_\mu)$ and the transformations~\eqref{eq:u1gal-local-transformations} is referred to as $U(1)$-Galilean geometry.

\subsubsection{SMT string Polyakov action and gauge fixing}
\label{sssec:smt-string-polyakov-action-gauge-fixing}
The SMT string sigma model coupling to a $U(1)$-Galilean geometry is given by~\cite{Harmark:2018cdl}
\begin{equation}
  \label{eq:smt-string-polyakov-action}
  S = - \frac{\tilde{T}}{2} \int d^2\sigma \left[
    2 \epsilon^{\alpha\beta} m_\alpha \pd_\beta \eta
    + e \theta^\alpha_1 \theta^\beta_1 h_{\alpha\beta}
    + \omega \epsilon^{\alpha\beta} e_\alpha^0 \tilde\tau_\beta
    + \psi \epsilon^{\alpha\beta} \left(e_\alpha^0 \pd_\beta \eta + e_\alpha^1 \tilde\tau_\beta\right)
  \right],
\end{equation}
where the tension $\tilde{T}$ is related to the relativistic string tension $T$ by $T = \tilde{T}/c$.
This is a Polyakov-type action, where the worldsheet geometry is described by a `zweibein' pair of one-forms $e^0$ and $e^1$ with corresponding dual vectors $\theta_0$ and $\theta_1$ that satisfy
\begin{equation}
  e^a_\alpha \theta^\alpha_b = \delta^a_b,
  \qquad
  e^a_\alpha \theta^\beta_a = \delta_\alpha^\beta.
\end{equation}
The worldsheet geometry is non-relativistic since $e^0$ and $e^1$ (and their dual vectors) play distinct roles in the action.
As a two-dimensional sigma model, the action~\eqref{eq:smt-string-polyakov-action} couples to the $U(1)$-Galilean geometry through the pullbacks
\begin{equation}
  \tilde\tau_\alpha = \pd_\alpha X^\mu \tilde\tau_\mu,
  \qquad
  m_\alpha = \pd_\alpha X^\mu m_\mu,
  \qquad
  h_{\alpha\beta} = \pd_\alpha X^\mu \pd_\alpha X^\nu h_{\mu\nu}.
\end{equation}
Due to the constraint associated to $\psi$, the string is forced to wind along a particular direction, which is described on the worldsheet by
\begin{equation}
  \label{eq:eta-winding-parametrization}
  \eta(\sigma^0,\sigma^1) = \frac{J}{2\pi T} \sigma^1 + \eta_\text{per}(\sigma^0,\sigma^1),
\end{equation}
where $\eta_\text{per}(\sigma^0,\sigma^1)$ is periodic in $\sigma^1\sim \sigma^1+2\pi$.
The winding parameter $J$ is identified with the $S^5$ or R-charge component of the charge~$Q$ in Equation~\eqref{eq:generic-charge-decomposition}.
It corresponds to the length of the spin chain in the large $N$ limit of the boundary Spin Matrix theory.
Finally, the action is invariant under the local transformations
\begin{equation}
  \label{eq:local-weyl-gal-boost}
  e^0 \to f e^0,
  \quad
  e^1 \to f e^1 + \hat{f} e^0,
  \quad
  \omega \to \omega / f - \hat{f} \psi / f^2,
  \quad
  \psi \to \psi/f.
\end{equation}
These are local Weyl transformations (parametrized by $f$) and local Galilean boosts (parametrized by $\hat{f}$).

\paragraph{Flat worldsheet gauge.}
In the following, we will mainly work with the action~\eqref{eq:smt-string-polyakov-action} in the flat worldsheet gauge, where we have
\begin{equation}
  \label{eq:smt-string-action-flat-gauge}
  e^0 = J d\sigma^0,
  \quad
  e^1 = d\sigma^1.
\end{equation}
In this gauge, the SMT string action~\eqref{eq:smt-string-polyakov-action} takes the form
\begin{equation}
  S_\text{flat} = - \frac{\tilde{T}}{2} \int d^2\sigma \left[
    2\epsilon^{\alpha\beta} m_\alpha \pd_\beta \eta
    + J h_{\mu\nu} X'^\mu X'^\nu
    + \omega J \tilde\tau_\mu X'^\mu + \psi (J \eta' - \tilde\tau_\mu \dot{X}^\mu)
  \right].
\end{equation}
Here and in the following, we denote $\sigma^0$-derivatives by dots and $\sigma^1$ derivatives by primes.
All $U(1)$-Galilean backgrounds that we obtain%
\footnote{
  Note that the alternative $u$-coordinate choice discussed in Footnote~\ref{fn:first-u-footnote} would lead to $-i\pd_u\to S+J$ instead of $-i\pd_u \to J$ in the SMT limit.
  This is less natural from the dual spin chain perspective, since the spin chain length is only counted by the total R-charge $J$.
  Additionally, it would result in backgrounds with $\tilde\tau=\cosh^2\rho d\tilde{x}^0$ instead of $\tilde\tau=d\tilde{x}^0$, which results in a more complicated gauge-fixed action.
}
in Section~\ref{sec:smt-string-backgrounds}
have $\tilde\tau = d\tilde{x}^0$.
With that, the Lagrange multipliers $\omega$ and $\psi$ imply
\begin{equation}
  (\tilde{x}^0)' = 0,
  \quad
  J \eta' = \dot{\tilde{x}}^0,
  \qiq
  \tilde{x}^0 = F(\sigma^0),
  \quad
  \eta = \frac{\dot{F}(\sigma^0)}{J} \sigma^1 + G(\sigma^0).
\end{equation}
The reparametrization freedom associated to $F(\sigma^0)$ and $G(\sigma^0)$ corresponds to a Galilean Conformal Algebra (GCA).
We can fix these residual symmetries by setting
\begin{equation}
  F(\sigma^0) = J^2 \sigma^0,
  \quad
  G(\sigma^0) = 0
  \qiq
  \tilde{x}^0 = J^2 \sigma^0,
  \quad
  \eta = J \sigma^1,
\end{equation}
so that the action becomes
\begin{equation}
  S_{\text{flat,gf}}
  = - J\tilde{T} \int d^2\sigma \left[
    m_\mu \dot{X}^\mu + \frac{1}{2} h_{\mu\nu} X'^\mu X'^\nu
  \right].
\end{equation}
In the following, we will mainly be interested in $U(1)$-Galilean backgrounds coming from AdS$_5\times S^5$, so that the resulting $m_\mu$ and $h_{\mu\nu}$ scale with the square of the AdS radius $R$.
Using $T = 1/(2\pi l_s^2) = \tilde{T}/c$ and $(R/l_s)^4=\lambda=4\pi g_s N$, the effective string tension is then
\begin{equation}
  T_\text{eff}
  = R^2 T = \frac{\sqrt{4\pi g_s N}}{2\pi} = \frac{1}{2\pi c},
  \qiq
  \tilde{T}_\text{eff} = \frac{1}{2\pi}.
\end{equation}
With that, the flat worldsheet gauge-fixed SMT string action is
\begin{equation}
  \label{eq:smt-flat-gf-string-action}
  S_{\text{flat,gf}}
  = - \frac{J}{2\pi} \int d^2\sigma \left[
    m_\mu \dot{X}^\mu + \frac{1}{2} h_{\mu\nu} X'^\mu X'^\nu
  \right].
\end{equation}
Here, we have absorbed the factors $R^2$ from $m_\mu$ and $h_{\mu\nu}$ in the effective string tension, which we will also do for the $U(1)$-Galilean backgrounds we obtain in the following.

\newpage
\section{SMT string backgrounds from \texorpdfstring{AdS$_5\times S^5$}{AdS5xS5}}
\label{sec:smt-string-backgrounds}
Ultimately, we want to construct a class of $U(1)$-Galilean backgrounds $(\tilde\tau_\mu, m_\mu, h_{\mu\nu})$ that are simple enough to allow us to quantize SMT strings on these backgrounds.
As can be seen from the gauge-fixed action~\eqref{eq:smt-flat-gf-string-action} the target space should clearly not be \emph{too} simple: if $m_\mu=0$, there is no dynamics left.

In the following, we will first obtain several $U(1)$-Galilean geometries that allow for non-trivial SMT string dynamics.
We start from AdS$_5\times S^5$ and work out the construction outlined in Section~\ref{ssec:smt-review-strings} for the SMT limits listed in Table~\ref{tab:smt-limits}.
The resulting backgrounds are curved and hence the corresponding sigma models are still rather complicated, but in Section~\ref{sec:penrose-limits} we will consider further simplifying limits of these backgrounds that lead to the class of backgrounds we are after.

While $U(1)$-Galilean string backgrounds were constructed for $SU(2)$ and $PSU(1,2|3)$ in earlier work~\cite{Harmark:2017rpg,Harmark:2018cdl}, we now provide a new systematic procedure and illustrate it by constructing the $SU(2|3)$ and $SU(1,1)$ backgrounds, which have not yet been considered.
Our procedure, which is outlined in Section~\ref{ssec:general-procedure-smt-string-backgrounds} below,
improves on previous constructions in two ways.
First, we consistently use $(x^0,u)$-coordinates in a way that leads to a convenient spin chain interpretation of the momentum along the $u$-direction.
Second, the sphere coordinates we use to derive the $PSU(1,2|3)$ background now allow us to obtain all other backgrounds as submanifolds of this geometry, mirroring how on the boundary all SMTs are contained in the $PSU(1,2|3)$ theory.

\subsection{General procedure}
\label{ssec:general-procedure-smt-string-backgrounds}
The recipe for constructing $U(1)$-Galilean geometries that was outlined in Section~\ref{ssec:smt-review-strings} can be implemented as follows.
First, choose one of the BPS bound $E\geq Q$ of $\mathcal{N}=4$ associated to the charges $Q$ listed in Table~\ref{tab:smt-limits}, which are of the form
\begin{equation}
  \label{eq:general-procedure-charge}
  Q
  = S + J.
\end{equation}
Such a charge is a particular combination of the commuting AdS$_5$ and $S^5$ angular momenta $(S_1,S_2)$ and $(J_1,J_2,J_3)$, respectively.
We parametrize AdS$_5\times S^5$ using
\begin{subequations}
  \label{eq:ads5-s5-embedding-coordinates}
  \begin{align}
    z_0
    &= R \cosh\rho\, e^{it},
    &
    w_1
    &= R \sin(\beta_1/2) \sin(\beta_2/2) e^{i\alpha_1},
    \\
    z_1
    &= R \sinh\rho\, \sin(\bar\beta/2) e^{i\bar\alpha_1},
    &
    w_2
    &= R \sin(\beta_1/2) \cos(\beta_2/2) e^{i\alpha_2},
    \\
    z_2
    &= R \sinh\rho\, \cos(\bar\beta/2) e^{i\bar\alpha_2},
    &
    w_3
    &= R \cos(\beta_1/2) e^{i\alpha_3}.
  \end{align}
\end{subequations}
Here and in the following, we use barred coordinates for angles on the $S^3\subset\text{AdS}_5$
and unbarred coordinates for angles on the $S^5$.
In this parametrization,
the commuting charges $S_i$ and $J_j$ correspond to the rotations
\begin{equation}
  S_i = - i \pd_{\bar\alpha_i},
  \quad
  J_j = - i \pd_{\alpha_j}.
\end{equation}
We then combine appropriate angles $\bar\alpha_i$ and $\alpha_j$ into new coordinates $\gamma$ and $\bar\gamma$ so that $S=-i\pd_{\bar\gamma}$ and $J=-i\pd_\gamma$.
Together with the AdS$_5$ global time $t$, we use these to define three coordinates $x^0$, $u$ and $w$ so that as per Equation~\eqref{eq:x0-u-vectors-charges} we have
\begin{equation}
  \label{eq:generic-smt-null-coordinates}
  \begin{pmatrix}
    t
    \\
    \bar\gamma
    \\
    \gamma
  \end{pmatrix}
  =
  \begin{pmatrix}
    1 & -1/2 & 0
    \\
    1 & -1/2 & c_1
    \\
    1 & 1/2 & c_2
  \end{pmatrix}
  \begin{pmatrix}
    x^0
    \\
    u
    \\
    w
  \end{pmatrix}
  \qiq
  \begin{cases}
  i \pd_{x^0} = E - S - J = E - Q,
  \\
  -i \pd_u = \frac{1}{2} \left(E - S + J\right),
  \\
  -i\pd_w
  = c_1 S + c_2 J.
  \end{cases}
\end{equation}
The parameters $c_1$ and $c_2$ control how the $w$-direction is aligned along
the $S^3\subset\text{AdS}_5$ and the $S^5$.
For the transformation to be invertible, we need $c_1\neq0$.
We keep these parameters arbitrary for now, but at the end of this section, we show that we can always gauge fix $c_1=1$ and $c_2=0$ so that $-i\pd_w = S$.
When we consider a SMT limit associated to one of the BPS bounds from Table~\ref{tab:smt-limits} that only involve $S^5$ momenta, such as the $SU(2|3)$ case in Section~\ref{ssec:su23-background}, we do not need the $\bar\gamma$ or $w$ coordinates.

Since (combinations of) the vectors $\pd_{\bar\alpha_i}$, $\pd_{\alpha_j}$ and $\pd_t$ are of constant length on particular submanifolds of AdS$_5\times S^5$, the vector $\pd_u$ will be null on a particular submanifold.
If we parametrize this submanifold $M$ using $(u,x^0,x^i)$ with $i=1,\ldots,2n$, we can therefore decompose its induced metric in terms of TNC variables as in Section~\ref{sssec:relation-to-tnc-strings},
\begin{equation}
  \label{eq:generic-submanifold-tnc-decomposition}
  \left. ds^2 \right|_M
    = 2 \tau_\mu dx^\mu (du - m_i dx^i) + h_{ij} dx^i dx^j,
  \qquad
  \tau = \tau_0 dx^0 + \tau_i dx^i.
\end{equation}
On the full background, the light-cone gauge Hamiltonian of the relativistic string contains a confining potential proportional to $1/g_s$ in all $8-2n$ directions transverse to $M$, so that in the SMT limit~\eqref{eq:smt-limit-strings} the dynamics of the string is restricted to this submanifold~\cite{Harmark:2008gm}.
Simultaneously, the coordinate $x^0$ is rescaled following~\eqref{eq:smt-limit-strings-using-c}.
Together, this leads to the $U(1)$-Galilean geometry%
\footnote{%
  Following the discussion in Section~\ref{ssec:smt-review-strings}, here and in the following we combine the factors of $R^2$ in the TNC variables $m_\mu$ and $h_{\mu\nu}$ with the string tension $T$ when taking the SMT limit, which leads to the effective rescaled tension $\tilde{T}_\text{eff}=1/2\pi$.
  For this reason, the $U(1)$-Galilean tensors $m_\mu$ and $h_{\mu\nu}$ are independent of the AdS radius $R$.
}
\begin{equation}
  \tilde\tau=\tau_0 d\tilde{x}^0,
  \quad
  h_{\mu\nu},
  \quad
  m_\mu,
\end{equation}
associated to the bulk dual of the Spin Matrix theory corresponding to $E\geq Q$.

\subsection{The \texorpdfstring{$SU(2|3)$}{SU(2|3)} background}
\label{ssec:su23-background}
As a first novel example, we study the $SU(2|3)$ theory, which has the largest possible compact spin group one can obtain from a Spin Matrix limit of $\mN=4$.
It corresponds to zooming in on the BPS bound given by $Q=J_1+J_2+J_3$, which involves all commuting $S^5$ generators from the bulk perspective.
As we will see, this leads to a $U(1)$-Galilean geometry with compact spatial sections corresponding to $\CC\PP^2$.
For this, we transform the $S^5$ angular coordinates $\alpha_j$ in the embedding coordinates~\eqref{eq:ads5-s5-embedding-coordinates} into three new coordinates $\chi,\psi,\vphi\in(0,2\pi)$ using
\begin{equation}
  \begin{pmatrix}
    \alpha_1
    \\
    \alpha_2
    \\
    \alpha_3
  \end{pmatrix}
  =
  \begin{pmatrix}
    1 & 1/2 & -1/2
    \\
    1 & 1/2 & 1/2
    \\
    1 & -1/2 & 0
  \end{pmatrix}
  \begin{pmatrix}
    \chi
    \\
    \psi
    \\
    \vphi
  \end{pmatrix}.
\end{equation}
As a result, we have
\begin{equation}
  Q = J_1 + J_2 + J_3 = -i\pd_\chi.
\end{equation}
After relabeling $\beta_1/2=\xi\in(0,\pi/2)$ and $\beta_2=\theta\in(0,\pi)$ in~\eqref{eq:ads5-s5-embedding-coordinates},
the induced metric is
\begin{equation}
  \label{eq:ads5-s5-metric-fubini-study-coordinates}
  ds^2 / R^2
  = - \cosh^2\rho\, dt^2 +  d\rho^2 + \sinh^2\rho d\bar\Omega_3^2
  +  \left(d\chi + B\right)^2 +  d\Sigma_2^2.
\end{equation}
Here, the $S^5$ is described as a circle fibration (parametrized by $\chi$) over a $\CC\PP^2$ base space,
in terms of the following potentials and Fubini--Study metrics,
\begin{equation}
  \label{eq:cp2-and-cp1-potentials-and-metric}
  \begin{gathered}
    B = \sin^2\xi \left(d\psi + A\right) - \frac{1}{2} d\psi,
    \quad
    A = \frac{1}{2} \cos\theta d\vphi,
    \\
    d\Sigma_2^2
    = d\xi^2 + \sin^2\xi d\Sigma_1^2
    + \sin^2\xi \cos^2\xi \left(d\psi + A\right)^2,
    \quad
    d\Sigma_1^2
    = \frac{1}{4} \left(d\theta^2 + \sin^2\theta d\vphi^2\right).
  \end{gathered}
\end{equation}
We can now introduce the coordinates $x^0$ and $u$ as in~\eqref{eq:generic-smt-null-coordinates} with $\gamma=\chi$,
\begin{equation}
  \begin{pmatrix}
    t
    \\
    \chi
  \end{pmatrix}
  =
  \begin{pmatrix}
    1 & -1/2
    \\
    1 & 1/2
  \end{pmatrix}
  \begin{pmatrix}
    x^0
    \\
    u
  \end{pmatrix}.
\end{equation}
Since $\pd_\chi$ is of constant length over the entire base $\CC\PP^2\subset S^5$, we have
\begin{equation}
  4 (\pd_u)^2 / R^2 = - \cosh^2\rho + 1 \leq 0,
\end{equation}
so that $u$ is null if and only if $\rho=0$.
The six-dimensional submanifold $M$ at $\rho=0$ is described by $x^0$, $u$ and the $\CC\PP^2$ coordinates $(\theta,\vphi,\xi,\psi)$.
The decomposition~\eqref{eq:generic-submanifold-tnc-decomposition} of the metric on $M$ then gives us the following TNC variables,
\begin{equation}
  \label{eq:su23-tnc-geometry}
  \tau = dx^0 + \frac{1}{2} B,
  \quad
  m = - R^2 B,
  \quad
  h = R^2 d\Sigma_2^2.
\end{equation}
In the SMT limit~\eqref{eq:smt-limit-strings}, the $x^0$-coordinate is rescaled, and $\tau$ is replaced by $\tilde\tau = d\tilde{x}^0$.
Absorbing the factors of $R^2$ in the effective string tension, the $U(1)$-Galilean geometry associated to the $SU(2|3)$ background is then
\begin{equation}
  \label{eq:su23-u1gal-geometry}
  \tilde\tau = d\tilde{x}^0,
  \quad
  m = - B,
  \quad
  h = d\Sigma_2^2,
\end{equation}
where the Fubini--Study $\CC\PP^2$ potential $B$ and metric $d\Sigma_2^2$ are given by~\eqref{eq:cp2-and-cp1-potentials-and-metric}.
Note that the $\CC\PP^2$ geometry includes an $S^3$, which we have parametrized as a Hopf fibration over a base $\CC\PP^1$.
This $\CC\PP^1$ corresponds to the $SU(2)\subset SU(2|3)$ background, which corresponds to zooming in on the BPS bound $E\geq Q=J_1+J_2$.
Setting $\xi=\pi/2$ and fixing $\psi$ reduces the $SU(2|3)$ geometry~\eqref{eq:su23-u1gal-geometry} to the known $SU(2)$ background~\cite{Harmark:2017rpg,Harmark:2018cdl}.
We will see a more general version of this pattern in Section~\ref{ssec:psu123-background-and-reductions}, where we show how the maximal $PSU(1,2|3)$ background can be reduced to obtain the $U(1)$-Galilean geometries associated to all SMT limits.

\subsection{The \texorpdfstring{$SU(1,1)$}{SU(1,1)} background}
\label{ssec:su11-background}
Next, we consider the $SU(1,1)$ limit, which corresponds to $Q=S_1+J_1$.
This is the simplest case that involves both $S^5$ and AdS$_5$ charges.
We start from the embedding coordinates~\eqref{eq:ads5-s5-embedding-coordinates}, where
the induced metric is
\begin{equation}
  \label{eq:ads5-s5-metric-cartan-coordinates}
  \begin{split}
    ds^2 / R^2
    &= - \cosh^2\rho\, dt^2 + d\rho^2 + \sinh^2\rho \left(
      \frac{d\bar\beta^2}{4}
      + \sin^2(\bar\beta/2) d\bar\alpha_1^2 + \cos^2(\bar\beta/2) d\bar\alpha_2^2
    \right)
    \\
    &{}\qquad
    +\frac{d\beta_1^2}{4} + \sin^2(\beta_1/2) \left(
      \frac{d\beta_2^2}{4} + \sin^2(\beta_2/2) d\alpha_1^2 + \cos^2(\beta_2/2) d\alpha_2^2
    \right)
    \\
    &{}\qquad
    + \cos^2(\beta_1/2) d\alpha_3^2.
  \end{split}
\end{equation}
In these coordinates, $S_1 = - i\pd_{\bar\alpha_1}$ and $J_1 = -i\pd_{\alpha_1}$.
Following~\eqref{eq:generic-smt-null-coordinates}, we then combine $\bar\alpha_1$ and $\alpha_1$ with the AdS time $t$ by introducing $x^0$, $u$ and $w$ using
\begin{equation}
  \label{eq:su11-orig-coord-redef}
  \begin{pmatrix}
    t
    \\
    \bar\alpha_1
    \\
    \alpha_1
  \end{pmatrix}
  =
  \begin{pmatrix}
    1 & -1/2 & 0
    \\
    1 & -1/2 & c_1
    \\
    1 & 1/2 & c_2
  \end{pmatrix}
  \begin{pmatrix}
    x^0
    \\
    u
    \\
    w
  \end{pmatrix}.
\end{equation}
We see that
\begin{equation}
  4(\pd_u)^2 / R^2
  = - \cosh^2\rho + \sinh^2\rho \sin^2(\bar\beta/2)
  + \sin^2(\beta_1/2) \sin^2(\beta_2/2)
  \leq 0,
\end{equation}
so $u$ is null if and only if
$\bar\beta=\beta_1=\beta_2=\pi$.
On this submanifold, we get the TNC data
\begin{subequations}
  \label{eq:su11-tnc-geometry}
  \begin{align}
    \tau
    &= dx^0 - \frac{1}{2} \left(c_1 \sinh^2\rho - c_2\right) dw,
    \\
    m / R^2
    &= - \left(c_1 \sinh^2\rho + c_2\right) dw,
    \\
    h / R^2
    &= d\rho^2
    + c_1^2 \sinh^2\rho \cosh^2\rho\, dw^2.
  \end{align}
\end{subequations}
In the Spin Matrix limit~\eqref{eq:smt-limit-strings}, the $x^0$-coordinate is rescaled and (absorbing $R^2$ in the effective string tension) we obtain
\begin{subequations}
  \label{eq:su11-u1gal-geometry}
  \begin{align}
    \tilde\tau
    &= d\tilde{x}^0,
    \\
    m
    &= - \left(c_1 \sinh^2\rho + c_2\right) dw,
    \\
    h
    &= d\rho^2
    + c_1^2 \sinh^2\rho \cosh^2\rho\, dw^2.
  \end{align}
\end{subequations}
Note that the spatial slices of this geometry (parametrized by $\rho$ and $w$) are not compact, in contrast to the $SU(2|3)$ geometry in~\eqref{eq:su23-u1gal-geometry}.
We will show at the end of this section that we can gauge fix $c_1=1$ and $c_2=0$.
The flat gauge-fixed action~\eqref{eq:smt-flat-gf-string-action} on this background is then given by
\begin{align}
  S_{\text{flat,gf}}
  &= - \frac{J}{2\pi} \int d^2\sigma \left[
    m_\mu \dot{X}^\mu + \frac{1}{2} h_{\mu\nu} X'^\mu X'^\nu
  \right]
  \label{eq:su11-flat-gf-string-action}
  \\
  &= \frac{J}{2\pi} \int d^2\sigma \left[
    \sin^2\rho\, \dot{w}
    - \frac{1}{2} \left(
      (\rho')^2 + \sinh^2\rho \cosh^2\rho (w')^2
    \right)
  \right].
\end{align}
With the correct coordinate identifications%
\footnote{%
  Note that our radial coordinate~$\rho$ corresponds to $\rho/2$ in~\cite{Bellucci:2004qr}.
}
this exactly reproduces the action obtained from coherent states in the $sl(2)$ spin chain and spinning strings on AdS$_5\times S^5$, see Equations~(3.11) and~(4.11) of~\cite{Bellucci:2004qr} and also~\cite{Stefanski:2004cw}.

\subsection{All backgrounds from the \texorpdfstring{$PSU(1,2|3)$}{PSU(1,2|3)} background}
\label{ssec:psu123-background-and-reductions}
Finally, we obtain the $U(1)$-Galilean geometry for $E\geq Q=S_1+S_2+J_1+J_2+J_3$.
This BPS bound leads to the $PSU(1,2|3)$ Spin Matrix theory, which can be restricted to obtain the Spin Matrix theories arising from all other bounds in Table~\ref{tab:smt-limits}.
Likewise, we show that the geometry we obtain here contains all other backgrounds as submanifolds.

For this, we use Hopf coordinates for the $S^3\subset\text{AdS}_5$ and we parametrize the $S^5$ using a $S^1$-fibration over $\CC\PP^2$ in Fubini--Study coordinates,
\begin{subequations}
  \label{eq:ads5-s5-hopf-fs-parametrization}
  \begin{align}
    z_0
    &= R \cosh\rho\, e^{it},
    &
    w_1
    &= R \sin\xi \sin(\theta/2) e^{i(\chi+\psi/2-\vphi/2)},
    \\
    z_1
    &= R \sinh\rho \sin(\bar\theta/2) e^{i( \bar\psi - \bar\vphi/2 )},
    &
    w_2
    &= R \sin\xi \cos(\theta/2) e^{i(\chi+\psi/2+\vphi/2)},
    \\
    z_2
    &= R \sinh\rho \cos(\bar\theta/2) e^{i( \bar\psi + \bar\vphi/2 )},
    &
    w_3
    &= R \cos\xi e^{i(\chi-\psi/2)}.
  \end{align}
\end{subequations}
As a result, we have $-i\pd_{\bar\psi}=S_1+S_2$ and $-i\pd_\chi=J_1+J_2+J_3$.
In terms of the Fubini--Study potentials and metrics in~\eqref{eq:cp2-and-cp1-potentials-and-metric}, the total metric is then given by
\begin{equation}
  \label{eq:psu123-coordinates-ads5-s5-metric}
  \begin{split}
    ds^2/R^2
    &= - \cosh^2\rho dt^2 + d\rho^2
    + \sinh^2\rho \left[
      d\bar\Sigma_1^2 + \left(d\bar\psi + \bar A\right)^2
    \right]
    + d\Sigma_2^2 + \left( d\chi + B \right)^2.
  \end{split}
\end{equation}
Following~\eqref{eq:generic-smt-null-coordinates}, we now define $(x^0,u,w)$ from the AdS time $t$ and the angles $\bar\psi$ and $\chi$,
\begin{equation}
  \label{eq:psu123-null-coordinates}
  \begin{pmatrix}
    t \\ \bar\psi \\ \chi
  \end{pmatrix}
  = \begin{pmatrix}
    1 & -1/2 & 0
    \\
    1 & -1/2 & c_1
    \\
    1 & 1/2 & c_2
  \end{pmatrix}
  \begin{pmatrix}
    x^0
    \\
    u
    \\
    w
  \end{pmatrix}.
\end{equation}
Note that
$\pd_{\bar\psi}$ and $\pd_\chi$ are of constant length, so that $\pd_u$ is now null on the entire AdS$_5\times S^5$ background.
As a result, we can rewrite the full metric~\eqref{eq:psu123-coordinates-ads5-s5-metric} in the form
\begin{equation}
  ds^2 = 2 \tau \left(R^2 du - m\right) + h,
\end{equation}
where the TNC variables are given by
\begin{subequations}
  \label{eq:psu123-tnc-geometry}
  \begin{align}
    \tau
    &= dx^0 - \frac{1}{2} \left(c_1 \sinh^2\rho - c_2\right) dw
    - \frac{1}{2} \sinh^2\rho \bar A
    + \frac{1}{2} B,
    \\
    m / R^2
    &= - \left(c_1 \sinh^2\rho + c_2\right) dw
    - \sinh^2\rho \bar A - B,
    \\
    h / R^2
    &= d\rho^2
    + \sinh^2\rho\, d\bar\Sigma_1^2
    + \sinh^2\rho \cosh^2\rho \left(c_1 dw + \bar{A}\right)^2
    + d\Sigma_2^2.
  \end{align}
\end{subequations}
After taking the SMT limit~\eqref{eq:smt-limit-strings}, this results in
\begin{subequations}
  \label{eq:psu123-u1gal-geometry}
  \begin{align}
    \tilde\tau
    &= d\tilde{x}^0,
    \\
    m
    &= - \left(c_1 \sinh^2\rho + c_2\right) dw
    - \sinh^2\rho \bar A - B,
    \\
    h
    &= d\rho^2
    + \sinh^2\rho\, d\bar\Sigma_1^2
    + \sinh^2\rho \cosh^2\rho \left(c_1 dw + \bar{A}\right)^2
    + d\Sigma_2^2,
  \end{align}
\end{subequations}
where the Fubini--Study potentials $(A,B)$ and metrics $(d\Sigma_1^2,d\Sigma_2^2)$ can be read off from Equation~\eqref{eq:cp2-and-cp1-potentials-and-metric}, with their barred versions given in terms of the coordinates $(\bar\theta,\bar\vphi,\bar\psi)$.
This is the $U(1)$-Galilean geometry associated to $PSU(1,2|3)$ Spin Matrix theory.

\paragraph{Gauging away $c_1$ and $c_2$.}
In the above, we have kept the parameters $c_1$ and $c_2$ arbitrary.
They are associated to the position of the $w$-circle in the AdS$_5$ and the $S^5$ factors.
In fact, it turns out that these parameter can be gauged away entirely for all the backgrounds considered above.
We can see this from the $PSU(1,2|3)$ TNC background in~\eqref{eq:psu123-tnc-geometry} using the following coordinate shift and $U(1)$ gauge transformation,
\begin{equation}
  \label{eq:gauging-away-c2}
  x^0 \to x^0 - \frac{1}{2} c_2 w,
  \qquad
  m \to m + d\sigma = m + c_2 dw.
\end{equation}
As a result, the coefficient $c_2$ is removed entirely from the TNC geometry~\eqref{eq:psu123-tnc-geometry}.
Removing $c_2$ from the $U(1)$-Galilean background~\eqref{eq:psu123-u1gal-geometry} after taking the SMT limit only requires the $U(1)$ gauge transformation above.
Without loss of generality, we can therefore assume from now on that $c_2=0$ and (rescaling $w$) we can also set $c_1=1$,
so that the $PSU(1,2|3)$ background in~\eqref{eq:psu123-u1gal-geometry} becomes
\begin{subequations}
  \label{eq:psu123-u1gal-geometry-no-c}
  \begin{align}
    \tilde\tau
    &= d\tilde{x}^0,
    \\
    m
    &= - \sinh^2\rho dw
    - \sinh^2\rho \bar A - B,
    \\
    h
    &= d\rho^2
    + \sinh^2\rho\, d\bar\Sigma_1^2
    + \sinh^2\rho \cosh^2\rho \left(dw + \bar{A}\right)^2
    + d\Sigma_2^2,
  \end{align}
\end{subequations}
On this background, the flat gauge-fixed SMT string action~\eqref{eq:smt-flat-gf-string-action} gives
\begin{align}
  S_{\text{flat,gf}}
  = \frac{J}{4\pi} \int d^2\sigma
  &\left[
    2\sinh^2\rho\, \dot{w}
    + \sinh^2\rho \cos\bar\theta\, \dot{\bar\vphi}
    - \cos(2\xi) \dot{\psi}
    + \sin^2\xi \cos\theta \, \dot{\vphi}
  \right.\nonumber
  \\
  &{}\quad\nonumber
  - (\rho')^2
  - \frac{1}{4} \sinh^2\rho
  \left( (\bar\theta')^2 + \sin^2\bar\theta (\bar\vphi')^2 \right)
  \\
  &{}\quad
  \label{eq:smt-flat-gf-string-action-psu123}
  - \sinh^2\rho \cosh^2\rho
  \left(
    w' + \frac{1}{2} \cos\bar\theta\, \bar\vphi'
  \right)^2
  \\
  &{}\quad\nonumber
  - (\xi')^2
  - \frac{1}{4} \sin^2\xi
  \left( (\theta')^2 + \sin^2\theta (\vphi')^2 \right)
  \\
  &{}\quad\nonumber
  \left.
    - \sin^2\xi \cos^2\xi
    \left(
      \psi' + \frac{1}{2} \cos\theta\, \vphi'
    \right)^2
  \right].
\end{align}
This action should correspond to the (bosonic part of the) sigma model one obtains from a limit of the action for the $PSU(1,2|3)$ spin chain in a coherent state representation.

\paragraph{Restrictions to subcases.}
In earlier work, two distinct forms of the $PSU(1,2|3)$ $U(1)$-Galilean backgrounds have been derived~\cite{Harmark:2017rpg,Harmark:2018cdl}.
These forms differ from the present result~\eqref{eq:psu123-u1gal-geometry-no-c} in the spherical parametrization used, as well as (in the case of~\cite{Harmark:2018cdl}) in the choice of $u$-coordinate.
As we discussed in Section~\ref{ssec:smt-review-strings}, the current choice of $u$-coordinate is preferable both for the dual spin chain interpretation as well as the form of the resulting flat gauge-fixed worldsheet action.
Furthermore, the sphere coordinates that we used above
give us a very convenient way to write down the $U(1)$-Galilean backgrounds for all SMT limits in Table~\ref{tab:smt-limits} as submanifolds of the $PSU(1,2|3)$ background~\eqref{eq:psu123-u1gal-geometry-no-c}.
The resulting backgrounds can also be obtained from a direct computation, as we did for $SU(2|3)$ and $SU(1,1)$ above, which leads to the same results as the restrictions below.

As we already observed at the end of Section~\ref{ssec:su23-background},
the Fubini--Study metric and potential~\eqref{eq:cp2-and-cp1-potentials-and-metric} on $\CC\PP^2$ reduce to those on $\CC\PP^1$ when $\xi=\pi/2$ and $\psi$ is fixed,
\begin{equation}
  \left.\left(d\Sigma_2^2, B\right)\right|_{\substack{\xi = \pi/2\\\psi\text{ fixed}}}
  = \left(d\Sigma_1^2, A\right).
\end{equation}
Likewise, we can reduce the $S^3\subset\text{AdS}_5$ to the $S_1$ circle by fixing the $\CC\PP^1$ coordinates $(\bar\theta,\bar\vphi)$, and we can get rid of the $S^3$ entirely by setting $\rho=0$.
Through appropriate combinations of these restrictions, we can obtain the $U(1)$-Galilean backgrounds for all integer SMT limits.
This procedure is summarized in Table~\ref{tab:psu123-background-restrictions} below.

\begin{table}[ht]
  \centering
  \begin{tabular}{r|c|c c c c c}
    & $n$
    & $\rho=0$
    & $\bar\theta,\bar\vphi$ fixed
    & $\xi=\pi/2$ \& $\psi$ fixed
    & $\theta,\vphi$ fixed
    \\
    \hline
    $SU(2)$
    &
    1
    & \checkmark & (\checkmark) & \checkmark & --
    \\
    $SU(2|3)$
    &
    2
    & \checkmark & (\checkmark) & -- & --
    \\
    $SU(1,1)$
    &
    1
    & -- & \checkmark & \checkmark & \checkmark
    \\
    $PSU(1,1|2)$
    &
    2
    & -- & \checkmark & \checkmark & --
    \\
    $SU(1,2|2)$
    &
    2
    & -- & -- & \checkmark & \checkmark
    \\
    $PSU(1,2|3)$
    &
    4
    & -- & -- & -- & --
  \end{tabular}
  \caption{Restrictions of the $PSU(1,2|3)$ background~\eqref{eq:psu123-u1gal-geometry-no-c} to subcases.}
  \label{tab:psu123-background-restrictions}
\end{table}

\section{Flat SMT string backgrounds and Penrose limits}
\label{sec:penrose-limits}
We have thus obtained the $U(1)$-Galilean backgrounds corresponding to all of the integer Spin Matrix limits.
However, the flat worldsheet gauge fixed SMT string action~\eqref{eq:smt-flat-gf-string-action}
\begin{equation}
  S_{\text{flat,gf}}
  = - \frac{J}{2\pi} \int d^2\sigma \left[
    m_\mu \dot{X}^\mu + \frac{1}{2} h_{\mu\nu} X'^\mu X'^\nu
  \right],
\end{equation}
is still rather complicated on these backgrounds.
To illustrate this, let us briefly look at the $SU(2)$ case, where the background and the resulting string action are
\begin{gather}
  \tilde\tau = d\tilde{x}^0,
  \qquad
  m = - A
  = - \frac{1}{2} \cos\theta d\vphi,
  \qquad
  h = d\Sigma_1^2
  = \frac{1}{4} \left(d\theta^2 + \sin^2\theta d\vphi^2\right),
  \\
  S_{\text{flat,gf}}
  = \frac{J}{2\pi} \oint d\sigma \left(
    \cos\theta \dot{\vphi} - \frac{1}{4} \left[(\theta')^2 + \sin^2\theta (\vphi')^2\right]
  \right).
\end{gather}
This is a non-linear, interacting theory, and although it can be quantized directly due to its integrability~\cite{Klose:2006dd}, it is useful to consider a simplifying limit.

For this, we take a large $J$ limit and simultaneously zoom in on excitations around the point $\theta = \pi/2$ and $\vphi=0$ by taking
\begin{equation}
  J \to \infty,
  \quad
  \theta = \frac{\pi}{2} + x/\sqrt{J},
  \quad
  \vphi = y/\sqrt{J},
  \quad
  x,y
  \text{ fixed}.
\end{equation}
This results in the following background and action,
\begin{gather}
  \label{eq:su2-large-charge-smt-data}
  \tilde\tau_0 = d\tilde{x}^0,
  \qquad
  m_0
  = \lim_{J\to\infty} J m
  = \frac{1}{2} x\, dy,
  \qquad
  h_0
  = \lim_{J\to\infty} J h
  = \frac{1}{4} \left(dx^2 + dy^2\right),
  \\
  S = \frac{1}{4\pi} \oint d\sigma \left(
    x \dot{y} - \frac{1}{2} \left[(x')^2 + (y')^2\right]
  \right).
\end{gather}
The result is a free theory which describes the free magnon limit of the $SU(2)$ sector.

The above is strongly reminiscent of the Penrose limit, where one zooms in on the geometry around a null geodesic.
Starting from AdS$_5\times S^5$ and zooming in on a null geodesic along the $S^5$ and AdS$_5$ factors, the resulting geometry is the ten-dimensional maximally supersymmetric pp-wave~\cite{Blau:2002dy}.
However, the coordinate expression of the pp-wave metric that one obtains depends on the precise null geodesic one starts out with.
Since the SMT limit is defined in terms of a particular coordinate pair $(x^0,u)$, the resulting $U(1)$-Galilean geometry that one obtains from the pp-wave in these coordinates will therefore be different.
Zooming in on a geodesic generated by $\pd_u$ at a point on the submanifold where this vector is null, one obtains a background of the form%
\footnote{%
  Similar Penrose limits of AdS$_5\times S^5$ that lead to pp-wave backgrounds with flat directions have been considered earlier in e.g.\ Refs.~\cite{Michelson:2002wa,Bertolini:2002nr,Grignani:2009ny}.
}
\begin{equation}
  \label{eq:general-n-flat-pp-wave}
  ds^2 / R^2
  = dx^0 \left(du + x_i d y^i  \right)
  + dx_i dx^i
  + dy_i dy^i
  + dx_a dx^a
  - x_a x^a (dx^0)^2,
\end{equation}
where $i=1,\ldots,n$ and $a=1,\ldots,8-2n$.

In the following, we will see that for each SMT coordinate system, the resulting pp-wave is of the form~\eqref{eq:general-n-flat-pp-wave} and has $2n$ `flat' transverse directions $(x^i, y^i)$.
The string action contains a quadratic potential for the $8-2n$ remaining transverse directions $x^a$.
The slope of this potential diverges in the SMT limit, since we rescale $x^0=\tilde{x}^0/c^2$, so the dynamics in these `curved' directions are suppressed, and only the `flat' directions remain.
This results in the following $U(1)$-Galilean backgrounds,
\begin{subequations}
  \begin{align}
    \tilde\tau
    &= d\tilde{x}^0,
    \\
    m
    &= - \sum_{i=1}^n x^i dy^i,
    \\
    h
    &= \sum_{i=1}^{n} \left((dx^i)^2 + (dy^i)^2\right).
  \end{align}
\end{subequations}
Since they contain the crucial `mass flux' terms $x^i dy^i$ which supports the dynamics in the flat directions $x_i$ and $y_i$, we refer to these $U(1)$-Galilean geometries as \emph{flat fluxed} (FF) backgrounds.
Note that one can slightly generalize these backgrounds to include a magnon mass parameter $\mu_i$ multiplying each component of $m$, but we will have $\mu_i=1$ in the following.
The number of flat directions for each SMT limit and their origins in AdS$_5\times S^5$ are summarized in Table~\ref{tab:flat-dirs}.

\begin{table}[ht]
  \centering
  \begin{tabular}{r| c | c | c | c | c }
    & $n$
    & $\rho, w\in\text{AdS}_5\times S^5$
    & $\bar\theta,\bar\vphi\in S^3\subset\text{AdS}_5$
    & $\theta, \vphi \in S^3\subset S^5$
    & $\xi, \psi \in S^5$
    \\ \hline
    $SU(2)$
    & 1
    & -- & -- & \checkmark & --
    \\
    $SU(2|3)$
    & 2
    & -- & -- & \checkmark & \checkmark
    \\
    $SU(1,1)$
    & 1
    & \checkmark & -- & -- & --
    \\
    $PSU(1,1|2)$
    & 2
    & \checkmark & -- & \checkmark & --
    \\
    $SU(1,2|2)$
    & 2
    & \checkmark & \checkmark & -- & --
    \\
    $PSU(1,2|3)$
    & 4
    & \checkmark & \checkmark & \checkmark & \checkmark
  \end{tabular}
  \caption{The $2n$ `flat' directions for each sector and their AdS$_5\times S^5$ origin.}
  \label{tab:flat-dirs}
\end{table}

Our goal in this section is to show that the SMT and the large $J$ or Penrose limits commute, and that they result in the same FF $U(1)$-Galilean geometry, as depicted in Figure~\ref{fig:smt-penrose-limits-diagram} on page~\pageref{fig:smt-penrose-limits-diagram}.
We will do this explicitly for $SU(2|3)$, $SU(1,1)$ and $PSU(1,2|3)$.
These examples illustrate the general procedure, and the other cases can be obtained in a similar way.
Following the discussion at the end of Section~\ref{ssec:psu123-background-and-reductions}, we gauge fix the parameters $c_1=1$ and $c_2=0$ in all geometries in the following.

\subsection{The \texorpdfstring{$SU(2|3)$}{SU(2|3)} flat background}
\label{ssec:su23-limits}
Let us first consider the $SU(2|3)$ case.
As we found in Section~\ref{ssec:su23-background}, the $U(1)$-Galilean background associated to this SMT is given by~\eqref{eq:su23-u1gal-geometry},
\begin{equation}
  \label{eq:su23-u1gal-geometry-repeat}
  \tilde\tau = d\tilde{x}^0,
  \quad
  m = - B,
  \quad
  h = d\Sigma_2^2.
\end{equation}
For the large charge limit of the corresponding sigma model, we zoom in on a point on the $\CC\PP^2$ by defining
\begin{equation}
  \theta = \frac{\pi}{2} + x/\sqrt{J},
  \quad
  \vphi = y / \sqrt{J},
  \quad
  \xi = \frac{\pi}{4} + q/\sqrt{J},
  \quad
  \psi = p / \sqrt{J}.
\end{equation}
Taking $J\to\infty$ as in~\eqref{eq:su2-large-charge-smt-data} then results in the flat geometry
\begin{subequations}
  \label{eq:su23-large-q-u1gal-data}
  \begin{align}
    \tilde\tau_0
    &= d\tilde{x}^0,
    \\
    m_0
    &= \frac{1}{4} x\, dy - q\, dp,
    \\
    h_0
    &= \frac{1}{8} \left(
      dx^2 + dy^2
    \right)
    + dq^2 + \frac{1}{4} dp^2.
  \end{align}
\end{subequations}

Next, we show that this geometry can be recovered from a SMT limit of the corresponding pp-wave.
To obtain this pp-wave, we write the full AdS$_5\times S^5$ metric in terms of the coordinates adapted to $SU(2|3)$ introduced in Section~\ref{ssec:su23-background},
We then zoom in on a null geodesic on the submanifold $M$ corresponding to $\rho=0$ by introducing
\begin{equation}
  \label{eq:su23-penrose-limit}
  R = R'/\epsilon,
  \quad
  u = U \epsilon^2,
  \quad
  \rho = r \epsilon,
  \quad
  \theta = \frac{\pi}{2} + x \epsilon,
  \quad
  \vphi = y \epsilon,
  \quad
  \xi = \frac{\pi}{4} + q \epsilon,
  \quad
  \psi = p \epsilon.
\end{equation}
In the Penrose limit $\epsilon\to0$, this results in
\begin{equation}
  \label{eq:su23-pp-metric}
  \begin{split}
    ds^2 / (R')^2
    &= 2\tau_0 ( dU - m_0) + h_0
    + dr^2 + r^2 d\bar\Omega_3^2
    - r^2 (dx^0)^2.
  \end{split}
\end{equation}
Here, the TNC variables describing $M$ are
\begin{subequations}
  \label{eq:su23-pp-tnc-data}
  \begin{align}
    \tau_0
    &= dx^0,
    \\
    m_0 / (R')^2
    &= \frac{1}{4} x\, dy - q\, dp,
    \\
    h_0 / (R')^2
    &= \frac{1}{8} \left(
      dx^2 + dy^2
    \right)
    + dq^2 + \frac{1}{4} dp^2.
  \end{align}
\end{subequations}
A relativistic string on the background~\eqref{eq:su23-pp-metric} experiences a quadratic potential $r^2$ in the four transverse directions parametrized by $dr^2 + r^2d\bar\Omega_3^2$.
In the SMT limit~\eqref{eq:smt-limit-strings}, we have $x^0=\tilde{x}^0/c^2$ with $c\to\infty$, and this potential becomes infinitely steep.
As a result, these four transverse directions are suppressed, and the dynamics of the string is restricted to the $U(1)$-Galilean geometry described by~\eqref{eq:su23-large-q-u1gal-data}.
Note that the radius $R'$ is absorbed in the effective string tension in the SMT limit.

\subsection{The \texorpdfstring{$SU(1,1)$}{SU(1,1)} flat background}
\label{ssec:su11-limits}
Next, let us turn to the $SU(1,1)$ case, where we found in Section~\ref{ssec:su11-background} that the relevant $U(1)$-Galilean geometry is given by
\begin{subequations}
  \label{eq:su11-u1gal-geometry-repeat}
  \begin{align}
    \tilde\tau
    &= d\tilde{x}^0,
    \\
    m
    &= - \sinh^2\rho\, dw,
    \\
    h
    &= d\rho^2
    + \sinh^2\rho \cosh^2\rho\, dw^2,
  \end{align}
\end{subequations}
Here, we can take a large charge limit using $\rho = r / \sqrt{J}$, which results in
\begin{subequations}
  \label{eq:su11-large-q-u1gal-data}
  \begin{align}
    \tilde\tau_0
    &= d\tilde{x}^0,
    \\
    m_0
    &= - r^2 dw,
    \\
    h_0
    &= dr^2 + r^2 dw^2.
  \end{align}
\end{subequations}
The corresponding Penrose limit can be obtained from the AdS$_5\times S^5$ coordinates in Section~\ref{ssec:su11-background} by introducing
\begin{equation}
  \label{eq:su11-penrose-limit}
  R = R'/\epsilon,
  \quad
  u = U \epsilon^2,
  \quad
  \rho = \epsilon r,
  \quad
  \beta_1 = \pi + b_1 \epsilon,
  \quad
  \beta_2 = \pi + b_2 \epsilon.
\end{equation}
In the limit $\epsilon\to 0$, the resulting geometry is
\begin{equation}
  \label{eq:su11-pp-metric}
  \begin{split}
    ds^2 / (R')^2
    &= 2 dx^0 \left( dU + r^2 \sin^2(\bar\beta/2) dw\right)
    \\
    &{}\qquad
    +dr^2 + r^2 \left(
      \frac{1}{4} d\bar\beta^2
      + \sin^2(\bar\beta/2) dw^2 + \cos^2(\bar\beta/2) d\bar\alpha_2^2
    \right)
    \\
    &{}\qquad
    + \frac{1}{4} \left(
      db_1^2 + b_1^2 d\alpha_3^2
      + db_2^2 + b_2^2 d\alpha_2^2
    \right)
    \\
    &{}\qquad
    - \left[
      r^2 \cos^2(\bar\beta/2) + \frac{1}{4} (b_1^2 + b_2^2)
    \right] (dx^0)^2.
  \end{split}
\end{equation}
Note that the second line of~\eqref{eq:su11-pp-metric} describes a flat $\RR^4$, which we can reparametrize as
\begin{align}
  \label{eq:su11-pp-two-plane-param}
    Z_1 = r \sin(\bar\beta/2) e^{iw}
    = r_1 e^{iw},
    \qquad
    Z_2 = r \cos(\bar\beta/2) e^{i\bar\alpha_2}
    = r_2 e^{i\bar\alpha_2}.
\end{align}
With this, we have
$r_1^2 = r^2 \sin^2(\bar\beta/2)$
and
$r_2^2 = r^2 \cos^2(\bar\beta/2)$,
so we can rewrite the pp-wave~\eqref{eq:su11-pp-metric} into the more familiar quadratic form
\begin{align}
  \label{eq:su11-pp-metric-rewritten}
    ds^2 / (R')^2
    &= 2 dx^0 \left( dU + m_0\right)
    + h_0
    + dr_2^2 + r_2^2 d\bar\alpha_2^2
    \\ \nonumber
    &{}\quad
    + \frac{1}{4} \left(
      db_1^2 + b_1^2 d\alpha_3^2
      + db_2^2 + b_2^2 d\alpha_2^2
    \right)
    - \left[
      r_2^2 + \frac{1}{4} (b_1^2 + b_2^2)
    \right] (dx^0)^2.
\end{align}
When we now take the SMT limit $c\to\infty$ with $x^0 = \tilde{x}^0/c^2$,
we see that we are restricted to the submanifold where $r_2=b_1=b_2=0$.
As we can see from~\eqref{eq:su11-pp-two-plane-param}, $r_2=0$ means $\bar\beta=\pi$ and $r=r_1$, so that we exactly reproduce the $U(1)$-Galilean data~\eqref{eq:su11-large-q-u1gal-data} that we derived from the large $J$ limit above.

\subsection{The \texorpdfstring{$PSU(1,2|3)$}{PSU(1,2|3)} flat background}
\label{ssec:psu123-limits}
Finally, let us look at $PSU(1,2|3)$, where we found in Section~\ref{ssec:psu123-background-and-reductions} that the relevant $U(1)$-Galilean geometry is
\begin{subequations}
  \label{eq:psu123-u1gal-geometry-repeat}
  \begin{align}
    \tilde\tau
    &= d\tilde{x}^0,
    \\
    m
    &= - \sinh^2\rho\, dw
    - \sinh^2\rho\, \bar A - B,
    \\
    h
    &= d\rho^2
    + \sinh^2\rho\, d\bar\Sigma_1^2
    + \sinh^2\rho \cosh^2\rho \left(dw + \bar{A}\right)^2
    + d\Sigma_2^2.
  \end{align}
\end{subequations}
For the large $J$ limit, we zoom in on $\rho=0$ and on a point on the $\CC\PP^2$,
\begin{equation}
  \label{eq:su112-large-q-limit}
  \rho = r / \sqrt{J},
  \quad
  \theta = \frac{\pi}{2} + x / \sqrt{J},
  \quad
  \vphi = y / \sqrt{J},
  \quad
  \xi = \frac{\pi}{4} + q / \sqrt{J},
  \quad
  \psi = p / \sqrt{J}.
\end{equation}
This leads to the geometry
\begin{subequations}
  \label{eq:psu123-large-q-snc-data}
  \begin{align}
    \tilde\tau_0
    &= d\tilde{x}^0,
    \\
    m_0
    &= - r^2 dw
    - r^2 \bar A
    + \frac{1}{4} x\, dy - q\, dp,
    \\
    h_0
    &= dr^2
    + r^2 d\bar\Sigma_1^2
    + r^2 \left(dw + \bar A\right)^2
    + \frac{1}{8} \left(
      dx^2 + dy^2
    \right)
    + dq^2 + \frac{1}{4} dp^2.
  \end{align}
\end{subequations}
Note that $h_0$ describes an eight-dimensional flat geometry, with four dimensions in a radial Hopf parametrization.
This geometry can equally be obtained via a Penrose limit.
Using the coordinates introduced in Section~\ref{ssec:psu123-background-and-reductions}, we define
\begin{equation}
  \label{eq:psu123-penrose-limit}
  R = R'/\epsilon,
  \quad
  u = U \epsilon^2,
  \quad
  \rho = r \epsilon,
  \quad
  \theta = \frac{\pi}{2} + x \epsilon,
  \quad
  \vphi = y \epsilon,
  \quad
  \xi = \frac{\pi}{4} + q \epsilon,
  \quad
  \psi = p \epsilon.
\end{equation}
In terms of the geometry~\eqref{eq:psu123-large-q-snc-data}, the limit~$\epsilon\to0$ of the metric~\eqref{eq:psu123-coordinates-ads5-s5-metric} is then
\begin{equation}
  \label{eq:psu123-pp-metric}
  \begin{split}
    ds^2 / (R')^2
    &= 2 \tau_0 \left(
      dU + m_0
    \right)
    + h_0.
  \end{split}
\end{equation}
All directions are preserved in the SMT limit, which only rescales
$\tau_0=dx^0\to\tilde\tau_0=d\tilde{x}^0$.

\section{Phase space interpretation of SMT string target space}
\label{sec:point-particle-limit}
So far, we have worked out the $U(1)$-Galilean SMT string backgrounds and their flat limits.
However, if we now consider the flat worldsheet gauge-fixed action~\eqref{eq:smt-flat-gf-string-action},
\begin{equation}
  \label{eq:smt-gauge-fixed-action-repeat}
  S = - \frac{J}{2\pi}  \int d\sigma^0 d\sigma^1 \left(
    m_\mu \dot{X}^\mu + \frac{1}{2} h_{\mu\nu} X'^\mu X'^\nu
  \right),
\end{equation}
we see that something remarkable has happened: in this gauge, the velocities of the target space embedding fields $X^\mu$ only appear linearly in the action.

\subsection{Phase space action and symplectic potential}
To examine what this implies consider the following.
We can write the canonical `phase space' form of a particle action in terms of $n$ position variables $q^\mu$ and their $n$ conjugate momenta $p_\mu$, but we may equally parametrize it using $2n$ generic coordinates $x^M$,
\begin{equation}
  \label{eq:general-phase-space-action-form}
  S
  = \int dt \left(
    p_\mu \dot{q}^\mu
    - H(p,q)
  \right)
  = \int dt \left(
    C_M(x) \dot{x}^M
    - H(x)
  \right).
\end{equation}
This gives us a one-form $C_M(x) dx^M$ on phase space, which is known as a \emph{symplectic potential} since its exterior derivative defines the symplectic form on the phase space,
\begin{equation}
  \frac{1}{2} \omega_{MN} dx^M \wedge dx^N = d \left(C_M dx^M\right).
\end{equation}
The Poisson bracket between the $x^M$ is then determined by the (pointwise) inverse of the $2n\times2n$ matrix $\omega_{MN}$,
\begin{equation}
  \{ x^M, x^N \} = \omega^{MN}.
\end{equation}
This result can be straightforwardly generalized to a field theory action.

Comparing the light-cone gauge SMT string action~\eqref{eq:smt-gauge-fixed-action-repeat} above to~\eqref{eq:general-phase-space-action-form}, we see that the $X^\mu$ embedding coordinates can now be viewed as \emph{phase space coordinates!}
On this phase space, the dynamics is given by the symplectic form and Hamiltonian
\begin{equation}
  \label{eq:phase-space-general-string-hamiltonian}
  \omega = - \frac{J}{2\pi} dm,
  \qquad
  H = \frac{J}{4\pi} \oint d\sigma^1 h_{\mu\nu} X'^\mu X'^\nu.
\end{equation}
For this, $dm$ needs to be non-degenerate on equal time slices, which is true for all the $U(1)$-Galilean geometries that we obtained in Sections~\ref{sec:smt-string-backgrounds} and~\ref{sec:penrose-limits}.

To illustrate how this arises due to the Spin Matrix limit, we carefully consider the limit of the Hamiltonian description of a point particle on the pp-wave background~\eqref{eq:general-n-flat-pp-wave} that we encountered in Section~\ref{sec:penrose-limits}.
This will lead to a trivial Hamiltonian
but we can still see the appearance of the symplectic form $\omega\sim~dm$ and the phase space interpretation of the embedding coordinates.
A careful treatment of the Hamiltonian analysis of the SMT string and its gauge fixing will be the subject of upcoming work~\cite{upcoming-hamiltonian-paper}.

\subsection{From a limit of the light-cone point particle Hamiltonian}
Our starting point is the worldline reparametrization-invariant point particle action
\begin{equation}
  \label{eq:general-point-particle-action}
  S = \frac{1}{2} \int d\lambda \left(
    e\inv g_{\mu\nu} \dot{x}^\mu \dot{x}^\nu - e m^2
  \right).
\end{equation}
We will consider a slight generalization of the pp-wave metric~\eqref{eq:general-n-flat-pp-wave},
\begin{equation}
  \label{eq:general-pp-wave-like-background}
  ds^2 = 2dx^0 \left(du - m_i dx^i\right) + \delta_{IJ} dx^I dx^J
  - \delta_{ab} x^a x^b (dx^0)^2.
\end{equation}
Here, we have split the eight transverse coordinates $x^I$ into $2n$ `flat' directions $x^i$ and $8-2n$ directions $x^a$ where the particle feels a quadratic potential.
We now construct the Hamiltonian associated to the action~\eqref{eq:general-point-particle-action}.
Then, we will discuss the appropriate gauge fixing and we will take the SMT limit.

On the background~\eqref{eq:general-pp-wave-like-background}, the action is
\begin{equation}
  \label{eq:pp-point-particle-action}
  S = \int d\lambda \left[
    e\inv \left(
      \dot{x}^0\dot{u} - \dot{x}^0 m_i \dot{x}^i
      + \frac{1}{2} \delta_{IJ} \dot{x}^I \dot{x}^J
      - \frac{1}{2} \delta_{ab} x^a x^b (\dot{x}^0)^2
    \right)
    - \frac{e}{2} m^2
  \right].
\end{equation}
The momenta associated to the coordinates $u$, $x^0$, $x^i$ and $x^a$ are
\begin{subequations}
  \label{eq:pp-ham-momenta}
  \begin{align}
    p_u
    &= e\inv \dot{x}^0,
    \\
    p_0
    &= e\inv \dot{u} - e\inv m_i \dot{x}^i
    - e\inv x^a x_a \dot{x}^0,
    \\
    p_i
    &= - e\inv m_i \dot{x}^0 + e\inv \dot{x}_i,
    \\
    p_a
    &= e\inv \dot{x}_a.
  \end{align}
\end{subequations}
With this, we obtain the following Hamiltonian,
\begin{align}
  H &= \dot{x}^0 p_0 + \dot{u} p_u + \dot{x}^i p_i + \dot{x}^a p_a - L
  \\
  \label{eq:pp-point-particle-ham}
  &= \frac{e}{2} \left[
    2 p_0 p_u + (p_i + p_u m_i) (p^i + p_u m^i)
    + p_a p^a
    + x^a x_a (p_u)^2
    + m^2
  \right].
\end{align}
The Hamiltonian consists only of a constraint (with the einbein $e$ as Lagrange multiplier) due to the worldline reparametrization symmetry of the point particle action~\eqref{eq:general-point-particle-action}.
Note that the $u$-momentum $p_u$ is conserved since the Hamiltonian is independent of $u$.

A convenient way to fix this reparametrization symmetry in the Hamiltonian formalism is as follows.
With the Hamiltonian~\eqref{eq:pp-point-particle-ham}, the phase space action is
\begin{equation*}
  S = \int d\lambda \left(
    \dot{x}^0 p_0 + \dot{u} p_u + \dot{x}^i p_i + \dot{x}^a p_a
    - H
  \right).
\end{equation*}
Now vary this action with respect to $e$ and $p_0$, and solve the resulting equations for
\begin{equation}
  e = \frac{\dot{x}^0}{p_u},
  \quad
  p_0 = - \frac{1}{2p_u} \left[
    (p_i + p_u m_i) (p^i + p_u m^i)
    + p_a p^a
    + x_a x^a (p_u)^2
    + m^2
  \right].
\end{equation}
This solves the constraint, but to fully fix its Lagrange multiplier $e$ we need to specify a parametrization $x^0(\lambda)$.
In analogy with the Spin Matrix limit, where $x^0=c^2\tilde{x}^0$, we now achieve the same scaling by setting $x^0 = c^2\lambda$.
With that, we obtain the following phase space action and light-cone gauge-fixed Hamiltonian,
\begin{subequations}
  \label{eq:pp-point-particle-LC-gauge-fixed-action-ham}
  \begin{gather}
    S_\text{gf}
    = \int d\lambda \left[
      \dot{u} p_u + \dot{x}^i p_i + \dot{x}^a p_a - H_\text{LC}
    \right],
    \\
    H_\text{LC}
    = - \dot{x}^0 p_0
    = \frac{c^2}{2p_u} \left[
      (p_i + p_u m_i) (p^i + p_u m^i)
      + p_a p^a
      + x_a x^a (p_u)^2
      + m^2
    \right].
  \end{gather}
\end{subequations}

We now want to understand the behavior of this system in the SMT limit, which corresponds to sending $c\to\infty$ while keeping the velocities and the conserved null momentum $p_u$ fixed.
Since we now have $e = c^2 p_u$, the momenta~\eqref{eq:pp-ham-momenta} are
\begin{subequations}
  \label{eq:pp-ham-momenta-gauge-fixed}
  \begin{align}
    p_a
    &= \frac{p_u \dot{x}_a}{c^2},
    \\
    p_i
    &= \frac{p_u}{c^2} \left(
      \dot{x}_i - c^2 m_i
    \right).
  \end{align}
\end{subequations}
In the limit $c\to\infty$, we therefore have $p_a\to0$ and $p_i\to -p_u m_i$, which reduces us to a subspace of the full phase space described by the action and the Hamiltonian in~\eqref{eq:pp-point-particle-LC-gauge-fixed-action-ham}.
To determine the corresponding reduced phase space, we introduce the constraints
\begin{equation}
  \Phi_a = p_a \approx 0,
  \qquad
  \Phi_i = p_i + p_u m_i.
\end{equation}
which implement the same reduction as the SMT limit.
Their time evolution under the light-cone Hamiltonian~\eqref{eq:pp-point-particle-LC-gauge-fixed-action-ham} is
\begin{align}
  \dot{\Phi}_a
  &= \{ \Phi_a, H_\text{LC} \}
  = - c^2 p_u x_a
  \qiq
  \chi_a = x_a \approx 0,
  \\
  \dot{\Phi}_i
  &= \{ \Phi_i, H_\text{LC} \}
  = \frac{c^2}{2p_u} \{ \Phi_i, \Phi_j \Phi^j \} \approx 0,
  \\
  \dot{\chi}_a
  &= \{ \chi_a, H_\text{LC} \}
  = \frac{c^2}{2p_u} \{ \chi_a, \Phi_b \Phi^b \}
  \approx 0.
\end{align}
So we get one set of secondary constraints $\chi_a$, which fixes the `curved' transverse directions $x^a$ to zero, corresponding to the quadratic potential in these directions becoming infinitely steep in the SMT limit.
On the reduced phase space $\Phi_a \approx \chi_a \approx \Phi_i \approx 0$,
\begin{equation}
  H_\text{LC} \approx \frac{c^2 m^2}{2 p_u}.
\end{equation}
Since this Hamiltonian is constant, there is no non-trivial dynamics on the reduced phase space.
For $H_\text{LC}$ to be finite as $c\to\infty$ we can set the particle mass $m$ to zero.
Note that our set of constraints is entirely second-class,
\begin{equation}
  \{ \chi_a, \Phi_b \} = \delta_{ab},
  \qquad
  \{ \Phi_i, \Phi_j \} = p_u ( \pd_j m_i - \pd_i m_j)
  = - p_u (dm)_{ij}.
\end{equation}
The Dirac bracket that follows from this set of second-class constraints vanishes on $x^a$ and $p_a$ (allowing us to consistently set them to zero) and satisfies
\begin{equation}
  \label{eq:pp-point-particle-limit-poisson-bracket}
  \{ x^i, x^j \}_\text{DB}
  = - \frac{1}{p_u} (dm)^{ij}.
\end{equation}
This shows that $-p_u dm$ provides the symplectic form on the reduced phase space arising from the SMT limit.

A more straightforward way of obtaining this result is by directly gauge fixing the Lagrangian form of the action~\eqref{eq:pp-point-particle-action}, with $x^0 = c^2 \lambda$ and $e = c^2 / p_u$,
which gives
\begin{equation}
  \label{eq:pp-vielbein-action-gauge-fixed}
  S_\text{gf} = \int d\lambda \left[
    \frac{p_u}{c^2} \left(
      c^2\dot{u} - c^2 m_i \dot{x}^i
      + \frac{1}{2} \delta_{IJ} \dot{x}^I \dot{x}^J
      - \frac{1}{2} c^4 \delta_{ab} x^a x^b
    \right)
    - \frac{c^2 m^2}{2p_u}
  \right].
\end{equation}
For this action to be finite in the $c\to\infty$ limit, we set $x^a=0$ and $m=0$.
With that,
\begin{equation}
  \lim_{c\to\infty} S_\text{gf}
  = \int d\lambda \left[
    p_u \dot{u} - p_u m_i \dot{x}^i
  \right].
\end{equation}
The first term describes the decoupled variable $u$ and its constant momentum $p_u$, while the second term leads to the symplectic form $\omega = -p_u dm$ and the Poisson bracket~\eqref{eq:pp-point-particle-limit-poisson-bracket}.

\section{Conclusions and outlook}%
\label{sec:outlook}
In this paper we have obtained a class of Spin Matrix theory (SMT) string backgrounds that we refer to as flat-fluxed (FF) backgrounds.
They are the simplest non-trivial backgrounds for the non-relativistic SMT string theories constructed in Refs.~\cite{Harmark:2017rpg,Harmark:2018cdl}.
These backgrounds are examples of $U(1)$-Galilean geometries, which resemble Newton--Cartan geometries but with a $U(1)$ gauge potential that does not transform under Galilean boosts.
Such geometries can be obtained by implementing a bulk version of the Spin Matrix limit for AdS$_5\times S^5$, and combining this limit with the Penrose or large charge limit leads to the FF backgrounds.

We have computed the $U(1)$-Galilean backgrounds associated to all integer SMT limits of $\mathcal{N}=4$ super-Yang--Mills that are listed in Table~\ref{tab:smt-limits}.
Furthermore, we have shown that the SMT limit and the large charge/Penrose limits leading to the FF backgrounds commute, as illustrated in Figure~\ref{fig:smt-penrose-limits-diagram}.
Finally, we have put forward a novel interpretation of the $U(1)$-Galilean backgrounds of the SMT string, which views the embedding fields as coordinates on a \emph{phase space}, with a symplectic form determined by the exterior derivative of the $U(1)$ potential.

The U(1)-Galilean background corresponding to the SMT limit with spin group $PSU(1,2|3)$ contains all other backgrounds (associated to subgroups of this spin group) as submanifolds.
Because of the relation between Spin Matrix limits~\cite{Harmark:2014mpa} and the spin chain limit considered by Kruczenski~\cite{Kruczenski:2003gt}, one can compare the SMT string sigma models to earlier literature.
In particular, as already remarked in~\cite{Harmark:2014mpa}, the SMT flat gauge-fixed sigma model for spin group $SU(2)$ is the well-known Landau-Lifshitz model, which  follows from the continuum limit of the $SU(2)$ spin chain.
The results of this paper now allow us to compare with the $SU(1,1)$ and $PSU(1,1|2)$ sigma models that were obtained in~\cite{Bellucci:2004qr,Stefanski:2004cw} and~\cite{Bellucci:2006bv} respectively, and using
the $U(1)$-Galilean backgrounds obtained in Section~\ref{sec:smt-string-backgrounds} we find perfect agreement.
Furthermore, the sigma model~\eqref{eq:psu123-u1gal-geometry-no-c} for $PSU(1,2|3)$ has not yet been obtained from a continuous limit of the corresponding spin chain.%
\footnote{%
  Recently, spinning string solutions have been obtained~\cite{Roychowdhury:2020dke} from the sigma model associated to the version of the $PSU(1,2|3)$ background that was derived in earlier work~\cite{Harmark:2018cdl}.
  It would be interesting to extend these results to the present background which is more natural from the spin chain perspective.
}

The phase space interpretation that arises from $U(1)$-Galilean geometries contributes to our understanding of SMT strings, but it may also be directly relevant to the resulting sigma models themselves.
The exact S-matrix of the Landau--Lifshitz model (corresponding to $SU(2)$ SMT) has been obtained using integrability by Klose and Zarembo~\cite{Klose:2006dd}.
There, a judicious field redefinition was used that makes it possible to quantize this theory on its curved background.
Using our phase space perspective, an equivalent field redefinition can be obtained from the Darboux coordinates on the corresponding symplectic manifold.
Generalizing this strategy to more complicated sigma models could potentially greatly simplify quantizing these models exactly.%
\footnote{%
  Note that the phase space interpretation also holds for sigma models on the curved $U(1)$-Galilean backgrounds in Section~\ref{sec:smt-string-backgrounds}, not just for the FF backgrounds in Section~\ref{sec:penrose-limits}.
  The symplectic form $\omega\sim dm$ that follows from the $PSU(1,2|3)$ background~\eqref{eq:psu123-u1gal-geometry-no-c} involves the natural symplectic forms $dA$ and $dB$ on the Kähler manifolds $\CC\PP^1$ and $\CC\PP^2$ that are contained in this background.
}

Furthermore, an alternative description of Spin Matrix theories with spin groups containing a non-compact $SU(1,1)$ factor has recently been developed in terms of simple lower-dimensional non-relativistic field theories~\cite{Harmark:2019zkn,Baiguera:2020jgy}.
Since these theories appear to be under good control even at strong coupling, it would be extremely interesting to establish a holographic relation to strings on the $SU(1,1)$ background that we derived here.

Regarding the SMT string sigma models themselves, many of the most pressing questions are related to their quantization.
The analysis of Section~\ref{sec:point-particle-limit} can be repeated starting from relativistic strings in light-cone gauge, which should produce the symplectic form $\omega\sim dm$ and the Hamiltonian~\eqref{eq:phase-space-general-string-hamiltonian} from the Spin Matrix limit~\eqref{eq:smt-limit-strings-using-c}.
After quantizing the resulting gauge-fixed theory, one should then check if the global symmetries of the $U(1)$-Galilean background are anomalous.
Next, one would want to covariantly quantize the SMT Polyakov action~\eqref{eq:smt-string-polyakov-action} on the FF backgrounds we obtained in Section~\ref{sec:penrose-limits}.
This retains the two-dimensional Galilean Conformal algebra (GCA) of reparametrization symmetries~\cite{Harmark:2018cdl} on the worldsheet, in parallel to the Virasoro symmetries of relativistic string theory.
Anomalies of the GCA symmetry could then provide criteria on consistent SMT string backgrounds.
Such criteria are presumably met by our FF backgrounds since they are derived from limits of a consistent background for the relativistic string, but they could for example explain why only particular numbers $n$ of flat directions are allowed.
We will report on some of these issues and related ones in upcoming work~\cite{upcoming-hamiltonian-paper}.

In addition, it would be very interesting to determine the beta functions of the SMT string sigma model
and to analyze the resulting low-energy effective action for the $U(1)$-Galilean geometry.
For the non-relativistic TNC and SNC string theories with relativistic worldsheets and associated Virasoro symmetries, beta functions have already been obtained~\cite{Gomis:2019zyu,Gallegos:2019icg,Yan:2019xsf}.
Again, the FF backgrounds we obtained here should be solutions to the resulting equations of motion, and it would be interesting to see if they are in some sense `maximally symmetric', in analogy to the pp-wave solution of IIB supergravity~\cite{Blau:2001ne}.

Finally, there are numerous other open issues to investigate in relation to the SMT string theories.
These include for example the role of supersymmetry in SMT strings, both for the worldsheet theory
as well as the non-relativistic target space.%
\footnote{%
  For the TNC string, see \cite{Blair:2019qwi} for a worldsheet supersymmetric version that is derived using the connection between non-relativistic strings and double field theory \cite{Morand:2017fnv}.
  See also the older works \cite{Gomis:2005pg} as well as \cite{Fontanella:2020eje} for recent work on non-relativistic integrable supersymmetric sigma-models obtained via the method of Lie algebra expansions.
}
Other directions include a  more detailed analysis of the allowed worldsheet geometries and the role of the dilaton coupling~\cite{Harmark:2019upf}, open string sectors, brane-like objects and the effect of the NS-NS $B$-field.

\subsection*{Acknowledgements}

We thank Lorenzo Menculini for early collaboration on this project and Arjun Bagchi, Stefano Baiguera, and Leo Bidussi for useful discussions.
The work of JH is supported by the Royal Society University Research Fellowship ``Non-Lorentzian Geometry in Holography'' (grant number UF160197).
The work of TH, NO and GO is supported in part by the project ``Towards a deeper understanding of  black holes with non-relativistic holography'' of the Independent Research Fund Denmark (grant number DFF-6108-00340) and the work of NO and GO also by the Villum Foundation Experiment project 00023086.

\bibliographystyle{JHEP}
\bibliography{smt-ff-backgrounds}

\end{document}